\documentclass[journal]{IEEEtran}

\usepackage{cite}
\usepackage{amsmath,amssymb,amsfonts}
\usepackage{algorithmic}
\usepackage{graphicx}
\usepackage{textcomp}
\def\BibTeX{{\rm B\kern-.05em{\sc i\kern-.025em b}\kern-.08em
		T\kern-.1667em\lower.7ex\hbox{E}\kern-.125emX}}

\usepackage{color}
\usepackage{bm}
\usepackage{algorithm}

\usepackage{hhline}

\ifCLASSOPTIONcompsoc
\usepackage[caption=false, font=normalsize, labelfont=sf, textfont=sf]{subfig}
\else
\usepackage[caption=false, font=footnotesize]{subfig}
\fi

\usepackage{soul}

\usepackage{babel}
\usepackage{threeparttable}

\begin{document}
	\title{Near-Field Millimeter-Wave Imaging via Arrays in the Shape of Polyline}
	\author{
		Shuoguang~Wang,
		Shiyong~Li,~\IEEEmembership{Member,~IEEE,}  
		Guoqiang~Zhao,
		and~Houjun~Sun 
		\thanks{
			The work was supported by the National Natural Science Foundation of China under Grant 61771049. (\textit{Corresponding author: Shiyong Li.})  }
		
		\thanks{The authors are with the Beijing Key Laboratory of Millimeter Wave and Terahertz Technology, Beijing Institute of Technology, Beijing 100081, China. (e-mail: lisy\_98@bit.edu.cn).}

	}
	
	\markboth{IEEE}
	{Shell \MakeLowercase{\textit{et al.}}: Near-Field Millimeter-Wave Imaging via Arrays in the Shape of Polyline}
	
	\maketitle

\begin{abstract}
		
This paper proposes a polyline shaped array based  system scheme, associated with mechanical scanning along the array's perpendicular direction, for near-field millimeter-wave (MMW) imaging. Each section of the polyline is a chord of a circle with equal length. The polyline array, which can be realized as a monostatic array or a multistatic one, is capable of providing more observation angles than the linear or planar arrays. Further, we present the related three-dimensional (3-D) imaging algorithms based on a hybrid processing in the time domain and the spatial frequency domain. The nonuniform fast Fourier transform (NUFFT) is utilized to  improve the computational efficiency. Numerical simulations and experimental results are provided to demonstrate the efficacy of the proposed method in comparison with the back-projection (BP) algorithm.
		
		
		
		
	\end{abstract}
	
	\begin{IEEEkeywords}
		Millimeter-wave (MMW) imaging, polyline shaped array,  near-field, time domain, spatial frequency domain.
	\end{IEEEkeywords}
	
	\IEEEpeerreviewmaketitle
	
	\section{Introduction}
	
	Millimeter-waves (MMWs) have the ability to penetrate regular barriers. Therefore, MMW imaging is currently being investigated for the detection of  concealed weapons or contraband carried by personnel \cite{sheen1,zhuge2,1d_mimo_cylindrical,mtt_TR_appr,tankai}. Also, MMW is nonionizing, so it has no health hazard at moderate power levels compared with X-rays \cite{sheen1}. 
	
	To reconstruct the three-dimensional (3-D) image of the target under test,   two-dimensional (2-D) antenna apertures are required, which are usually formed by the one-dimensional (1-D) monostatic array accompanied by mechanical scanning.
	The monostatic array needs to satisfy the Nyquist sampling criterion to avoid image aliasing, which results in high number of antennas and switches. 
	One way to counter this drawback is to utilize the multiple-input multiple-output (MIMO) arrays which combine all the possible transmit-receive antenna pairs to  obtain more equivalent phase centers than the monostatic scenario \cite{Kirchhoff,zhuge2,ahmed1,subsampled_array}. 
	
	Various imaging schemes employing 1-D \cite{zhuge1,1dmimo_2, 1dmimo, 1d_mimo_cylindrical,guangyou} or 2-D MIMO arrays \cite{qps,zhuge_ap,zhuge2,2dmimo,tankai2,nufft_mimo2d} have been researched with different antenna configurations. 
	The 2-D MIMO array systems are able to obtain real-time imaging due to eliminating the mechanical scanning. However, the 2-D apertures could also result in  much higher system cost and complexity than those achieved by 1-D antenna arrays with mechanical scanning. 
	Therefore, this paper focuses only on the latter. 
	
	For the 1-D array scheme,  antennas are usually placed along a straight line, associated with a mechanical scanning either along a straight track  \cite{zhuge1,1dmimo_2, 1dmimo, guangyou} or a circular one  \cite{1dmimo_2, 1d_mimo_cylindrical}. 
	Furthermore, the typical 1-D MIMO array topology is configured by two separate dense transmit elements or subarrays set at both ends of the undersampled receive array. 
	Consequently, this type of arrays cannot provide an even illumination of a long target area along the array direction, however, which is  exactly the case for imaging of human body. 
	A single-frequency MIMO-arc array-based azimuth imaging approach was presented in \cite{mimo_arc_imaging}, based on the geometry transformation from the arc array to the corresponding equivalent linear array. 
	A circular-arc MIMO array scheme  as well as a wavenumber domain imaging algorithm was proposed in \cite{mimo_arc_imaging0},  for near-field 3-D imaging.  Combining the mechanical scanning along the array's perpendicular direction, all the antenna beams of the circular-arc array can be steered towards the target to be imaged. 
	
	In this paper, we present an array topology in the shape of polyline, each section of which lies in a chord of a cirle with equal length. 
	The transmit and receive antennas are evenly placed along the polyline. The antennas can be configured as either a monostatic array or a multistatic one. The array is mechanically scanned along its perpendicular direction to form an aperture that consists of several rectangular planes, as illustrated in Fig. \ref{array_geometry}. 
	Compared with the circular-arc array, the polyline one is much easier to be fabricated, which meanwhile can	keep the antenna beams almost pointing at the target center in the horizontal plane.
	
	We propose the imaging algorithms for both the monostatic and multistatic polyline array-based systems for the sake of completeness. 
	The  algorithms are formulated through the hybrid processing in the time domain and the spatial frequency domain.
	Specifically, we process the data in the spatial frequency domain for the mechanical scanning direction, while applying the time domain processing of the data along the array direction.
	Moreover, to improve the efficiency of the time domain processing, we apply the inverse nonuniform fast Fourier transform (NUFFT) \cite{nufft} to replace the conventional interpolations and the inverse fast Fourier transform (IFFT).
	
	
	The rest of the paper is organized as follows: In the next section, we formulate the 3-D imaging algorithms for the generic monostatic and multistatic polyline array-based systems. The inter-element spacings and resolutions are outlined. The  performance of the proposed imaging algorithms is demonstrated by simulations and experimental results in Section III. Finally, conclusions follow at the end.

	\section{MMW Imaging via Polyline Arrays}
	
	\subsection{Monostatic Polyline Array-Based Imaging}

	\begin{figure}[!t]
		\centering
		\subfloat[]{\label{a}
			\includegraphics[width=2.2in]{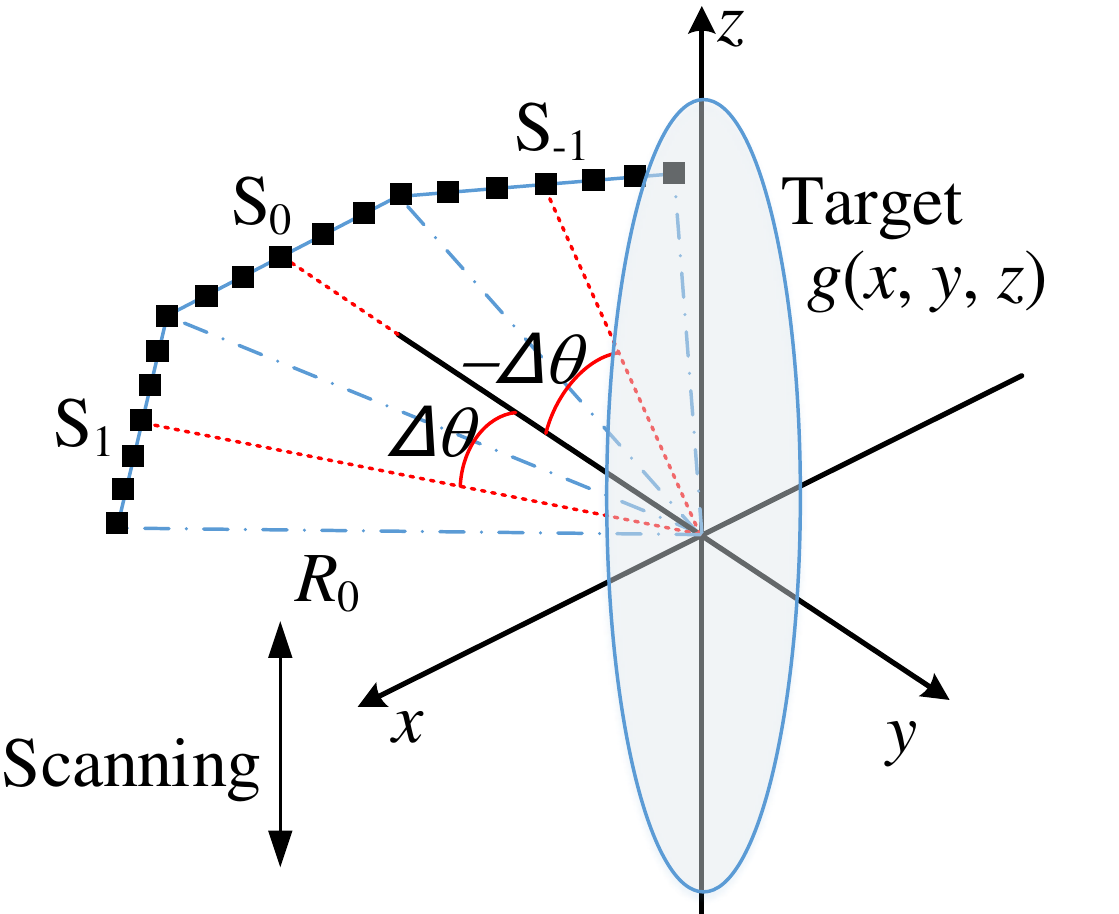}}
		\hfill
		\subfloat[]{\label{b}
			\includegraphics[width=2.4in]{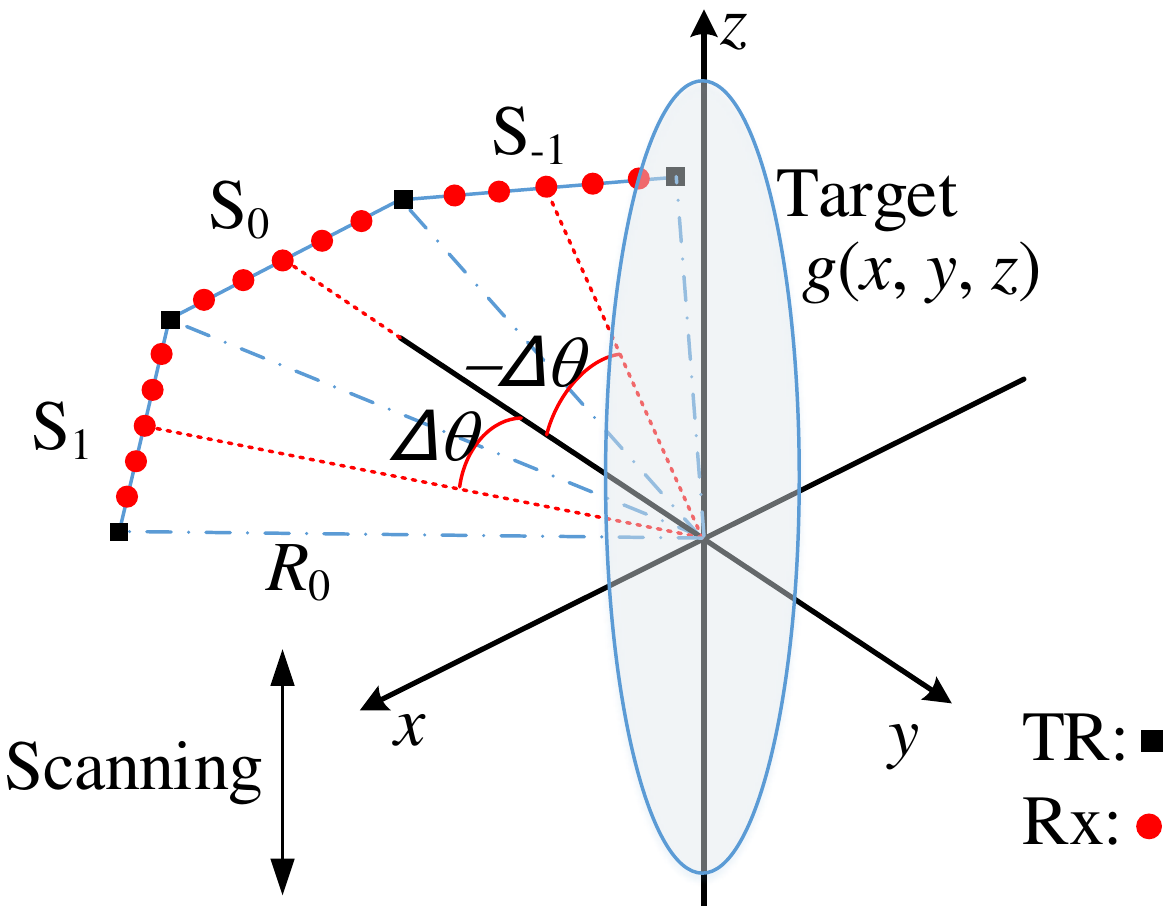}}
		\hfill
		\\	
		\caption{Topology of MMW imaging based on the polyline (a) monostatic array, and (b) multistatic array, respectively.}
		\label{array_geometry}
	\end{figure}
	

	The imaging geometry based on the monostatic array in the shape of polyline is shown in Fig. \ref{array_geometry}\subref{a}, where the transmit-receive (TR) antenna pairs are uniformly spaced along a polyline, meanwhile, scanning along the vertical direction. 
	The demodulated scattered waves are given by, 
	\begin{equation}\label{scat_wave1}
		s(x',y',z',k)\!=\!\iiint \!\! g(x,y,z)e^{-\mathrm{j}2kR}\mathrm{d}x\mathrm{d}y\mathrm{d}z, 
	\end{equation}
	where $k=\frac{2\pi f}{c}$ denotes the wavenumber, $f$ is the working frequency, $c$ is the speed of light, $g(x,y,z)$ represents the scattering coefficient of the target located at $(x,y,z)$, $R$ is the distance between the antenna and the target pixel, which is given by,
	\begin{equation}\nonumber
		R=\sqrt{(x-x')^2+(y-y')^2+(z-z')^2}, 
	\end{equation}
	where $(x',y',z')$ represents the position of the TR antennas.
	
	Next, we derive an imaging algorithm from this polyline array-based scheme based on the hybrid processing in the time domain and the spatial frequency domain. 
	
	
	We perform the Fourier transform on both sides of \eqref{scat_wave1} with respect to $z'$, then obtain, 
	\begin{align}\label{scat_wave_kz}
		&s(x',y',k_{z'},k)=\iiint g(x,y,z) \mathcal{F}_{z'}[e^{-\mathrm{j}2kR}]\mathrm{d}x\mathrm{d}y\mathrm{d}z. 
	\end{align}
	The Fourier transform of the free space Green's function $e^{-\mathrm{j}k2R}$ is expressed as \cite{soumekh}, 
	\begin{equation} \label{gr0}
		\mathcal{F}_{z'}[e^{-\mathrm{j}k2R}]= e^{-\mathrm{j} k_{\rho}\rho}e^{-\mathrm{j}k_{z'}z},
	\end{equation}
	where 
	\begin{equation}
		\rho=\sqrt{(x-x')^2+(y-y')^2},
	\end{equation}
	and 
	\begin{equation} \label{krho1}
		k_{\rho}=\sqrt{4k^2-k^2_{z'}}.
	\end{equation}
	
	Substituting \eqref{gr0} in \eqref{scat_wave_kz} yields,
	\begin{align}\label{scat_wave_kz0}
		s(x',y',k_{z},k)=\iiint g(x,y,k_{z})e^{-jk_\rho\rho}\mathrm{d}x\mathrm{d}y,
	\end{align}
	where the primed coordinate $k_{z'}$ is dropped to $k_z$  since the coordinate systems coincide. 
	
	Then, based on the time domain processing, we obtain,
	\begin{equation} \label{bp0}
		g(x,y,k_{z})= \iiint s(x',y',k_{z},k)e^{jk_\rho\rho}k\mathrm{d}k\mathrm{d}x'\mathrm{d}y',
	\end{equation}
	which can be solved by the back-projection algorithm, according to the following two steps, 
	\begin{subequations}
		\begin{align} 
			&q(l,k_z) =\int s(x',y',k_{z},k)e^{jk_\rho' l}k\mathrm{d}k,\label{mcbp0} \\ 
			&g(x,y,k_z) = \iint q(\rho,k_z)e^{jk_{\rho_0} \rho} \text{d}x'\text{d}y', \label{mcbp1}
		\end{align}
	\end{subequations}
	where $k_\rho' = k_{\rho}-k_{\rho_0}$, $k_{\rho_0} = \sqrt{4k_0^2-k_z^2}$ with $k_0$ denoted as the minimum value of $k$.

	The procedures indicated in \eqref{mcbp0} and \eqref{mcbp1} are referred to as the convolution (or filtered) back-projection algorithm \cite{cbp}. 
	Eq. \eqref{mcbp0} represents an inverse Fourier transform from $(k_\rho')$ to $l$. 
	Since $k$ is uniformly sampled, while  $k_\rho'$ is not, the inverse fast Fourier transform (IFFT) can not be directly applied.  
	Here, we utilize the inverse nonuniform fast Fourier transform (NUFFT) \cite{nufft} to improve the computational efficiency of \eqref{mcbp0}.
	Afterwards, the term 
	$q(\rho,k_z)$ in \eqref{mcbp1} is obtained through the 1-D interpolation of $q(l,k_z)$. 
	
	
	Finally, the image $g(x,y,z)$ can be obtained by,
	\begin{equation}
		g(x,y,z)=\text{IFFT}_{k_z}g(x,y,k_z).
	\end{equation}

	\subsection{Multistatic Polyline Array-Based Imaging}
	
	The imaging scenario via the polyline multistatic array is illustrated in Fig. \ref{array_geometry}\subref{b}. The transmit and receive antennas are still uniformly spaced along the polyline, with coordinates denoted by $(\vec{r}_T, z')$ and $(\vec{r}_R, z')$, respectively, where  $\vec{r}_T=(x_T',y_T')$ and $\vec{r}_R=(x_R',y_R')$.  However, the transmit subarray is undersampled (vice versa). Here, we only configure the transmit antennas at the ends of each polyline section.  
	The scattered waves after demodulation are expressed as,
	\begin{equation}\label{scat_wave2}
		s(\vec{r}_T,\vec{r}_R,z',k)\!=\!\iiint \!\! g(x,y,z)e^{-\mathrm{j}k(R_T+R_R)}\mathrm{d}x\mathrm{d}y\mathrm{d}z, 
	\end{equation}
	where
	\begin{align}\nonumber
		R_T &=\sqrt{\rho_T^2+(z-z')^2},  \\ 
		R_R &=\sqrt{\rho_R^2+(z-z')^2},   \nonumber
	\end{align}
	with 
	\begin{subequations}
		\begin{align}
			\rho_T=\sqrt{(x-x_T')^2+(y-y_T')^2}, \label{rhoT} \\
			\rho_R=\sqrt{(x-x_R')^2+(y-y_R')^2}. \label{rhoR}
		\end{align}
	\end{subequations}
	
	Similar to the processing of monostatic array, we first perform the Fourier transform over $z'$ to transform the data along the vertical dimension into the related spatial frequency domain, such as,
	\begin{align}\label{scat_wave_kz00}
		&s(\vec{r}_T,\vec{r}_R,k_{z'},k)=\iiint g(x,y,z)\cdot \\ \nonumber
		&\mathcal{F}_{z'}[e^{-\mathrm{j}kR_T}]\circledast_{k_{z'}} \mathcal{F}_{z'}[e^{-\mathrm{j}kR_R}]\mathrm{d}x\mathrm{d}y\mathrm{d}z. 
	\end{align}
	However, this is different from the monostatic case due to the convolution of the Green's functions in the Fourier domain. Similar to \eqref{gr0}, the Fourier transform of the Green's function can be written as,
	\begin{subequations}
		\begin{align} 
			\mathcal{F}_{z'}[e^{-\mathrm{j}kR_T}]= e^{-\mathrm{j}\sqrt{k^2-k^2_{z'}}\rho_T}e^{-\mathrm{j}k_{z'}z}, \label{grT} \\
			\mathcal{F}_{z'}[e^{-\mathrm{j}kR_R}]= e^{-\mathrm{j}\sqrt{k^2-k^2_{z'}}\rho_R}e^{-\mathrm{j}k_{z'}z}. \label{grR}
		\end{align}
	\end{subequations}
	
	Substituting \eqref{grT} and \eqref{grR} in \eqref{scat_wave_kz00} and using the property of convolution, we obtain,
	\begin{align}\label{scat_wave_kz2}
		&s(\vec{r}_T,\vec{r}_R,k_{z'},k)=\iiint g(x,y,z)e^{-\mathrm{j}k_{z'}z}\cdot \\ \nonumber
		&[e^{-\mathrm{j}\sqrt{k^2-k^2_{z'}}\rho_T}\circledast_{k_{z'}} e^{-\mathrm{j}\sqrt{k^2-k^2_{z'}}\rho_R}]\mathrm{d}x\mathrm{d}y\mathrm{d}z. 
	\end{align}
	The convolution in the square brackets is hard to be solved. But, we can find an approximated analytical solution through employing the approximation of $\rho_T\approx \rho_R$ (denoted by $\rho_0$) in the deducing procedure.
	Therefore, the convolution is given by,
	\begin{align}\label{conv_krho}
		&e^{-\text{j}\sqrt{k^2-k^2_{z'}}\rho_T}\circledast_{k_{z'}} e^{-\text{j} \sqrt{k^2-k^2_{z'}}\rho_R} \\ \nonumber
		&\approx e^{-\text{j}\sqrt{k^2-k^2_{z'}}\rho_0}\circledast_{k_{z'}} e^{-\text{j}\sqrt{k^2-k^2_{z'}}\rho_0} \\ \nonumber
		&=\int e^{-\text{j}\sqrt{k^2-\zeta^2}\rho_0}e^{-\text{j}\sqrt{k^2-(k_{z'}-\zeta)^2}\rho_0}\mathrm{d}\zeta.
	\end{align}
	Using the principle of stationary phase (POSP) \cite{cumming}, we obtain,
	\begin{align}\label{conv_krho_1}
		&e^{-\text{j}\sqrt{k^2-k^2_{z'}}\rho_T}\circledast_{k_{z'}} e^{-\text{j} \sqrt{k^2-k^2_{z'}}\rho_R}\\ \nonumber 
		&\approx e^{-\text{j}\sqrt{k^2-\frac{k_{z'}^2}{4}}\rho_T}e^{-\text{j}\sqrt{k^2-\frac{k_{z'}^2}{4}}\rho_R},
	\end{align}
	where the envelope and the constant terms are omitted. 
	Note that  $\rho_0$ is replaced by the original $\rho_T$ and $\rho_R$ after the convolution. 
	This handling  was first mentioned in \cite{mimo_arc_imaging0} to deal with the circular-arc MIMO imaging in the spatial-frequency domain. 
	The error introduced by the approximation is quite small \cite{mimo_arc_imaging0}.

	Substituting \eqref{conv_krho_1} in \eqref{scat_wave_kz2} yields, 
	\begin{align}\label{scat_wave_kz3}
		s(\vec{r}_T,\vec{r}_R,k_{z},k)=\iiint g(x,y,k_z)e^{-jk_\rho(\rho_T+\rho_R)}\mathrm{d}x\mathrm{d}y,
	\end{align}
	where 
	\begin{equation} \label{dispersion}
		k_\rho = \sqrt{k^2-\frac{k_{z}^2}{4}}.
	\end{equation}
	
	Clearly, the implementation of \eqref{scat_wave_kz3} can then use the same way as that of the monostatic array in Section II. A, which is given by,
	\begin{subequations}
		\begin{align} 
			&q(l,k_z) =\int s(\vec{r}_T,\vec{r}_R,k_{z},k)e^{jk_\rho' l}k\mathrm{d}k,\label{cbp0} \\ 
			&g(x,y,k_z) = \iint q(\rho_T+\rho_R,k_z)e^{jk_{\rho_0}(\rho_T+\rho_R)} \text{d}\vec{r}_T\text{d}\vec{r}_R, \label{cbp1}
		\end{align}
	\end{subequations}
	where  $k_\rho' = k_{\rho}-k_{\rho_0}$, $k_{\rho_0} = \sqrt{k_0^2-k_z^2/4}$. 
	Similarly, \eqref{cbp0} can be calculated by using the inverse NUFFT.  
	
	Then, the 3-D image $g(x,y,z)$ can be acheived by applying an IFFT of $g(x,y,k_z)$ with respect to $k_z$. 
	
	\subsection{Sampling Criteria}
	
	\begin{figure}[!t]
		\centering
			\includegraphics[width=2.0in]{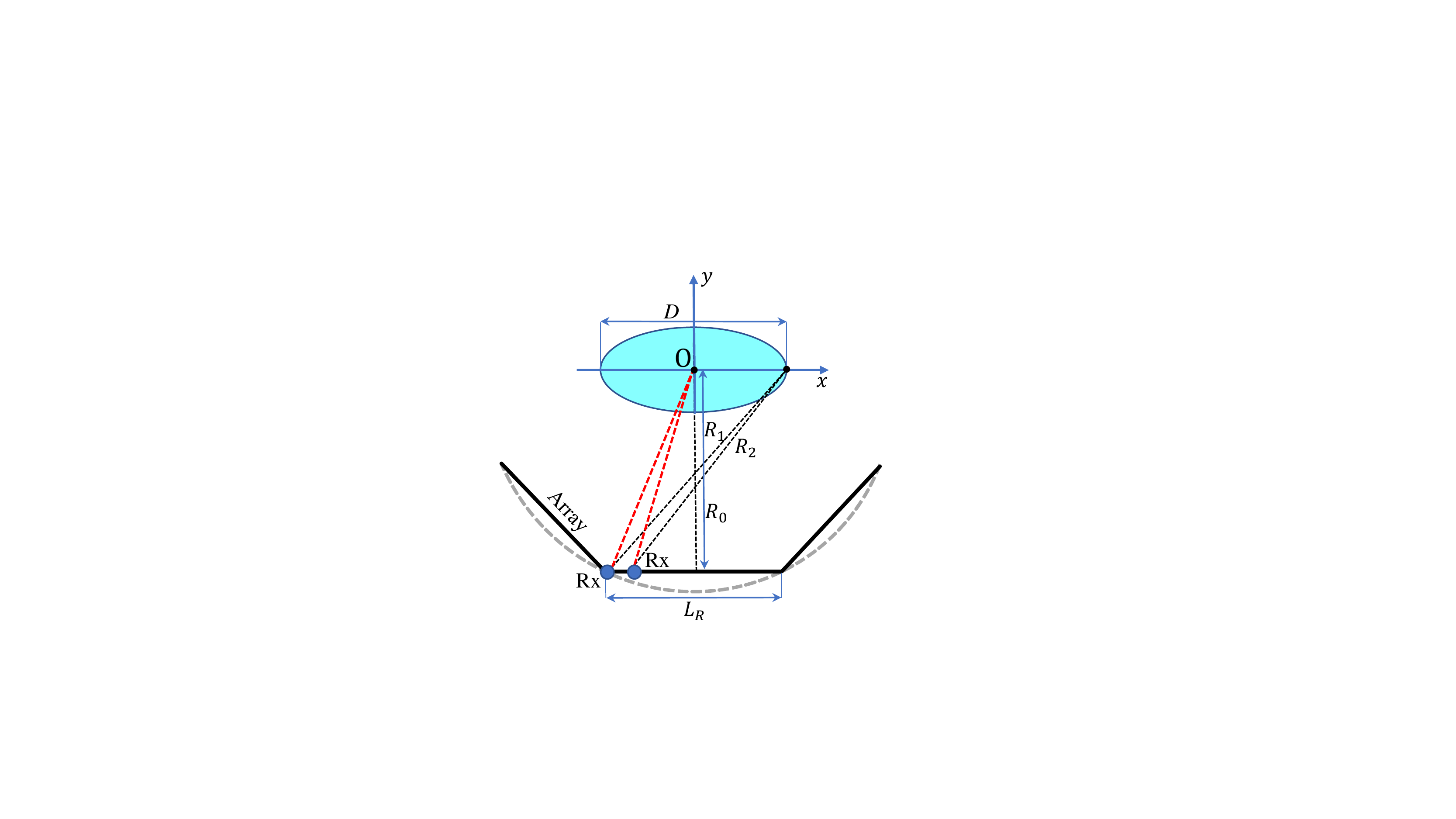}
		\caption{Illustration of the sampling criterion of polyline  array.}
		\label{sampling_circular}
	\end{figure}

	Considering the multistatic case, we configure that the transmit array is undersampled, and the receive array is fully sampled. 
	Firstly, the requirement for the inter-element spacing of the fully sampled receive array is presented. To avoid the aliasing effect in imaging results, the phase difference between two adjacent receive antennas has to be less than $\pi$ rad \cite{zhuge2}. According to the illustration in Fig. \ref{sampling_circular}, we can use the middle subarray without loss of generality to deduce  the requirement for antenna spacing, 
	\begin{equation}
		k(R_1-R_2)\approx k \Delta d \frac{(L_R+D)/2}{\sqrt{(L_R+D)^2/4+R_0^2}}\leq \pi,
	\end{equation}
	where $\Delta d$ denotes the spacing of two neighbouring receive antennas, $D$ and $R_0$ represent the maximal target dimension and the minimal distance from the middle subarray to the target center, respectively, in the horizontal plane, $L_R$ denotes the length of the subarray. 
	Then, we obtain,
	\begin{equation} \label{interval_mimo}
		\Delta d\leq  \frac{\lambda_{\text{min}}\sqrt{(L_R+D)^2/4+R_0^2}}{L_R+D},
	\end{equation}
	where $\lambda_{\text{min}}$ denotes the minimum wavelength of the working EM waves. Clearly, it is consistent with the requirement for the traditional linear MIMO array. 
	
	As for the undersampled transmit array, there is no restriction on the antenna spacing due to the fact that the image is obtained through multiplication between the transmit and receive array patterns. 
	In this work, we put transmit antennas at the joints and ends of each polyline section. 
	

	With regard to the monostatic case, the inter-element spacing of antennas is just half of \eqref{interval_mimo} due to the two-way EM wave propagation.

	The requirement for the mechanical scanning step along the vertical direction should also satisfy the phase difference limitation $k_{z_{\text{max}}}\Delta z' \leq \pi$ to avoid aliasing. Based on  \eqref{dispersion}, we have $k_{z_{\text{max}}} = 2k_{\text{max}}\sin\frac{\Theta_z}{2}$, then, the sampling constraint along the vertical aperture is given by,
	
	\begin{equation}
		\Delta z' \leq \frac{\lambda_{\text{min}}}{4\sin\frac{\Theta_z}{2}},
	\end{equation}
	where $\Theta_z$ denotes the minimal one between the antenna beamwidth along the scanning direction and the angle subtended by the scanning length from the center of the imaging scene.

	\subsection{Resolutions}
	
	First, the cross-range resolution along the direction of the polyline array is analyzed. Note that the cross-range resolution of either monostatic or multistatic arrays is determined by the extent of the corresponding spatial frequency, which leads to, 
	\begin{equation}
		\delta x = \frac{\pi}{k_{x_{\text{max}}}},
	\end{equation}
	where $k_{x_{\text{max}}}$ denotes the maximum of spatial frequency $k_x$ with respect to the horizontal direction $x$.
	
	According to  Fig. \ref{array_geometry}, $k_x$ can be considered a component of $k_{\rho}$ in \eqref{krho1} for the monostatic scenario or  in \eqref{dispersion} for the multistatic case, i.e., $k_x = k_{\rho}\sin \theta$ for the former, and $k_x = 2k_{\rho}\sin \theta$ for the latter, since  $k_x$ can be expressed as addition of $k_{x_T}$ and $k_{x_R}$, which have a same extent for the multistatic configuration in Fig. \ref{array_geometry}\subref{b}.
	Then, based on \eqref{krho1} and \eqref{dispersion}, we have $k_{x_{\text{max}}}\approx 2k_c\sin(\Theta_h /2)$, where $k_c$ denotes the center wavenumber of the working EM waves, and $\Theta_h$ represents the  angle subtended by the circumscribed arc of the polyline array, assuming that the  beamwidth of each antenna can fully illuminate the target in the horizontal direction.
	Thus, we have
	\begin{equation}
		\delta x \approx \frac{\pi}{2k_c\sin \frac{\Theta_h}{2}} = \frac{\lambda_c}{4\sin\frac{\Theta_h}{2}},
	\end{equation}
	for both the monostatic and multistatic scenarios.
	
	In the similar way, the resolution along the $z$ direction is given by,
	\begin{equation}
		\delta z =\frac{\lambda_c}{4\sin\frac{\Theta_z}{2}},
	\end{equation}
	where $\Theta_z$ represents the minimal one between the angle subtended by the scanning length and the antenna beamwidth.

	Finally, the down-range resolution is expressed as 
	\begin{equation}
		\delta y = \frac{c}{2B},
	\end{equation}
	where $B$ represents the bandwidth of the working waves.

	\section{Results}
	
	\begin{table}
		\centering
		\caption{Simulation Parameters}
		\setlength{\tabcolsep}{3pt}
		\begin{threeparttable}
			\begin{tabular}{p{200pt}  p{30pt}}
				\hline\hline
				Parameters& Values \\[0.5ex]
				\hline
				Radius of the circumscribed arc of the array $(R_0)$&
				1.0 m\\[0.5ex]
				Start frequency& 
				30 GHz \\[0.5ex]
				Stop frequency&
				35 GHz \\[0.5ex]
				Number of frequency steps&
				51 \\[0.5ex]
				Scanning step along elevation&
				0.5 cm \\[0.5ex]
				Number of polyline sections&
				3 \\[0.5ex]
				Number of transceiver antenna pairs per polyline section (monostatic)&
				40 \\[0.5ex]
				Spacing of transceiver antenna pairs along each polyline section (monostatic)&
				0.48 cm \\[0.5ex]
				Total number of transmit antennas  (multistatic)&
				4 \\[0.5ex]
				Number of receive antennas per polyline section (multistatic)&
				20 \\[0.5ex]			
				Spacing of transmit antennas (multistatic)&
				19.4 cm \\[0.5ex]
				Spacing of receive antennas along each polyline section (multistatic)&
				1.02 cm \\[0.5ex]
				
				\hline
			\end{tabular}
		\end{threeparttable}
		\label{tab1}
	\end{table}

	This section presents the simulation and experimental  results of the proposed method in comparison with BP. 
	The simulation paramters are listed in Table \ref{tab1} for both the monostatic and multistatic schemes. 
	Note that the antenna spacing of the monostatic array is set to be about half of  the multistatic case (specifically, referring to the spacing of the receive antennas) according to the discussion of sampling criteria in Section II. C.

	\begin{figure}[!t]
		\centering
		\subfloat[]{\label{a}
			\includegraphics[width=1.68in]{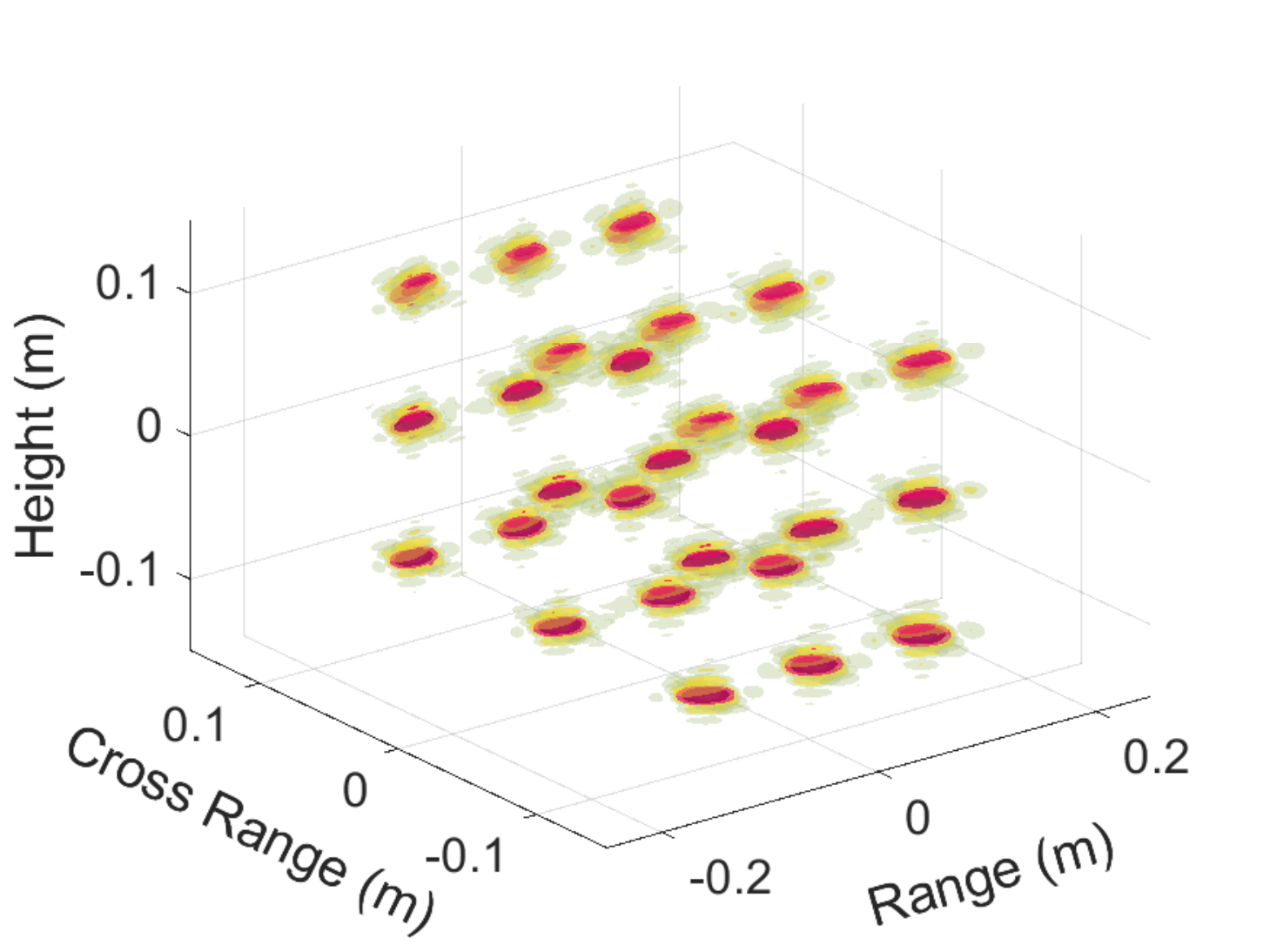}}
		\hfill
		\subfloat[]{\label{b}
			\includegraphics[width=1.68in]{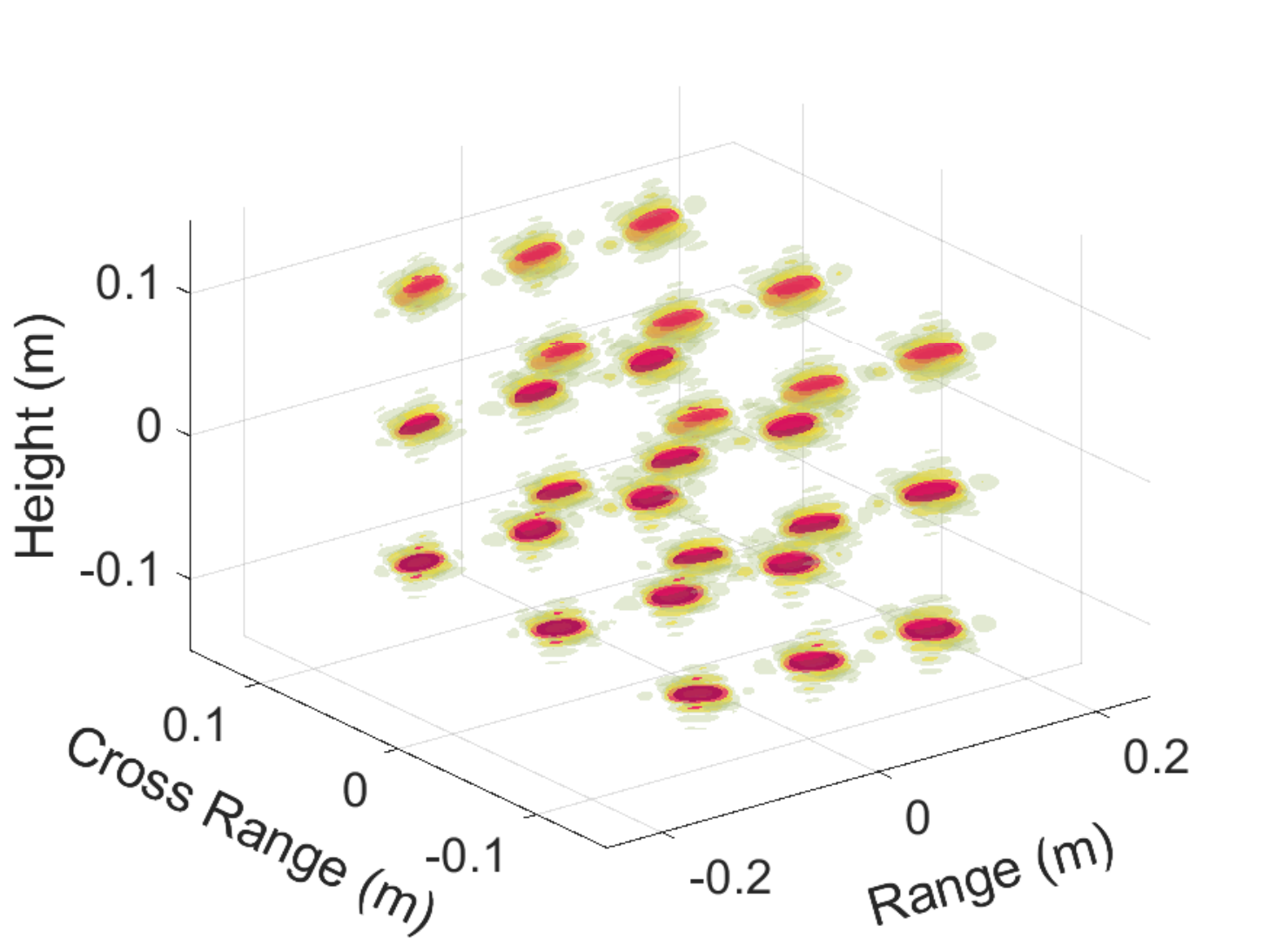}}
		\hfill
		\centering
		\subfloat[]{\label{c}
			\includegraphics[width=1.68in]{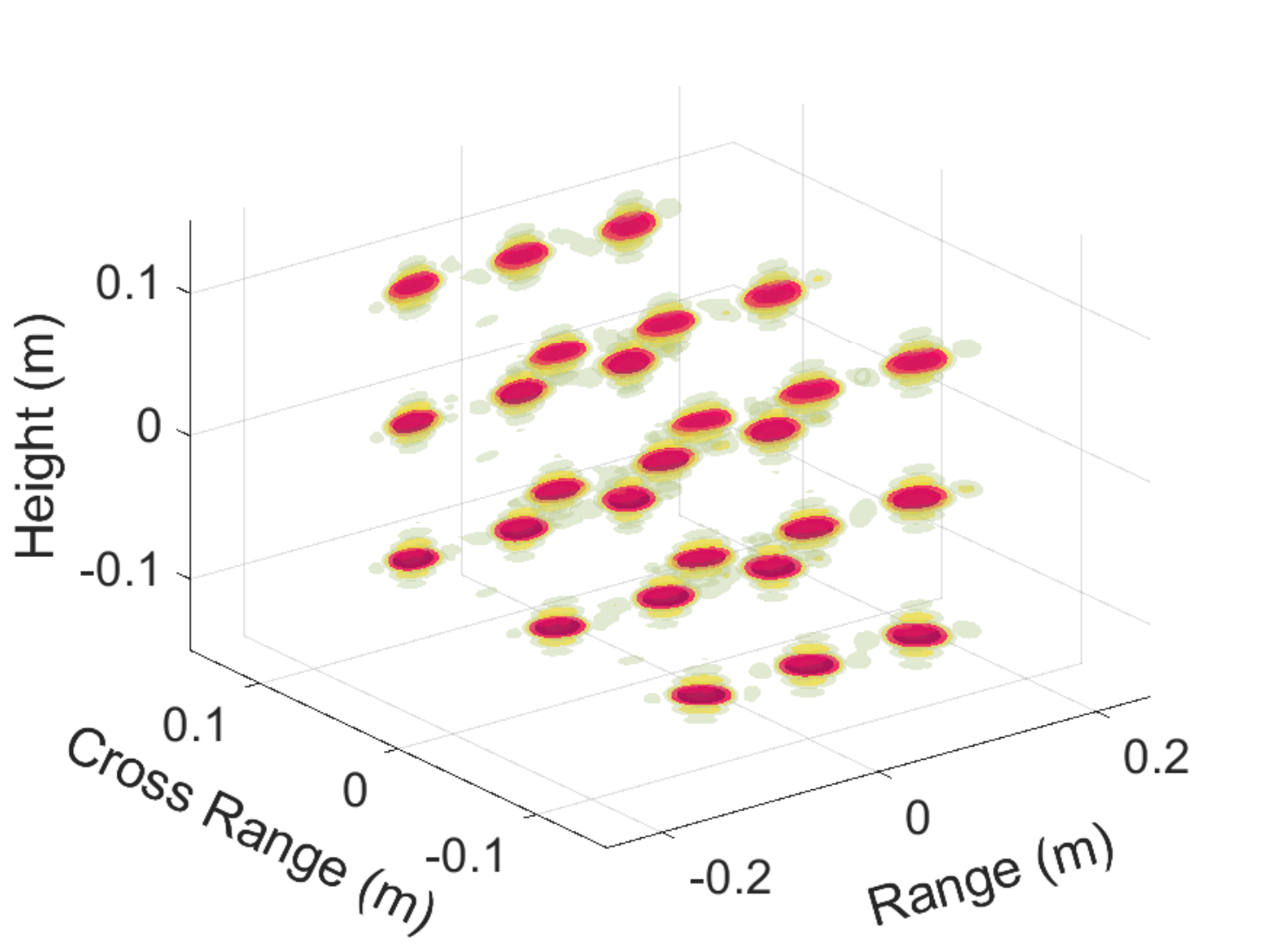}}
		\hfill
		\subfloat[]{\label{d}
			\includegraphics[width=1.68in]{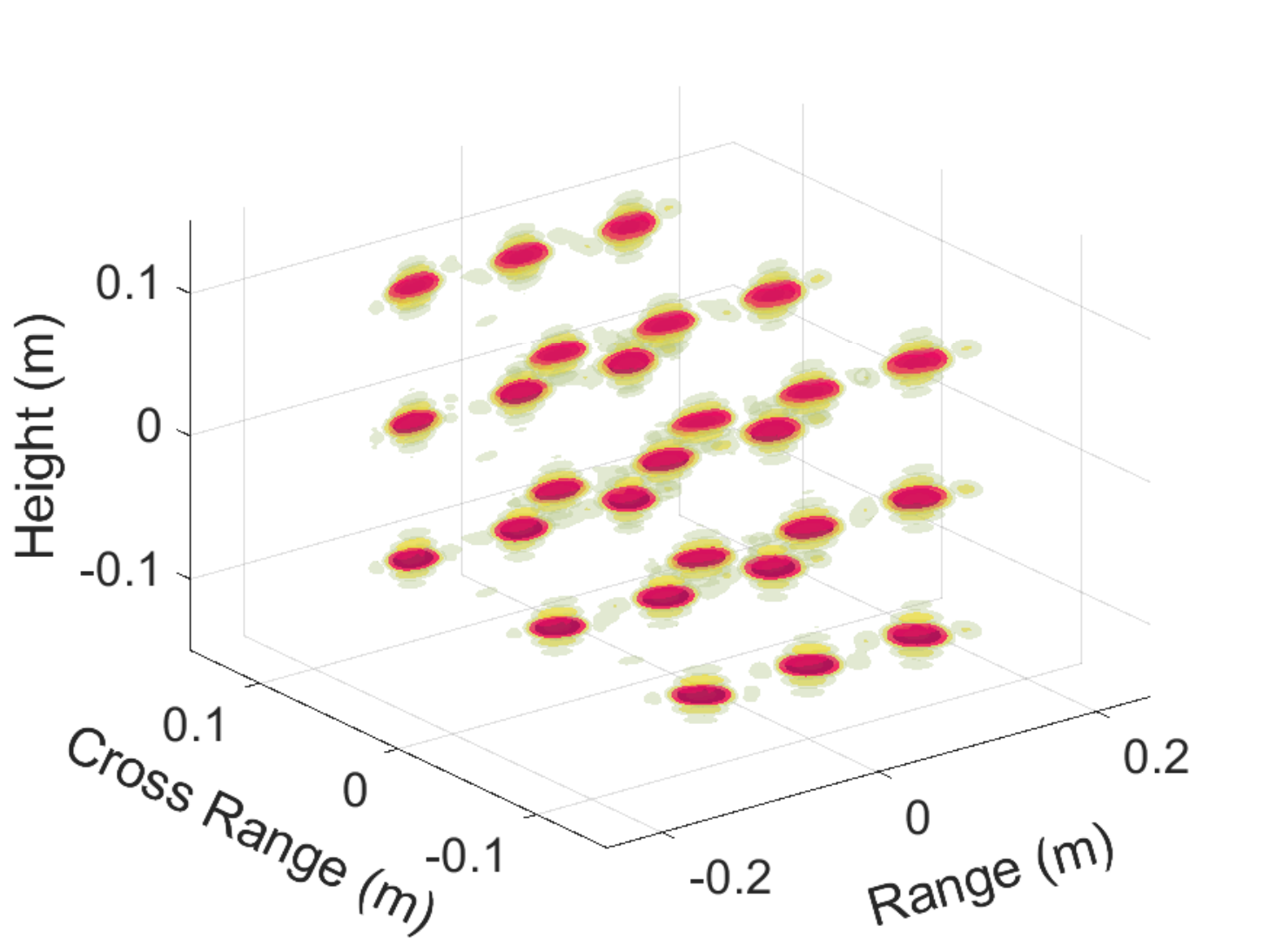}}
		\hfill
		\\	
		\caption{3-D imaging results via monostatic array: (a) the proposed algorithm, (b) BP; and multistatic array: (c) the proposed algorithm, (d) BP, respectively, with
			a dynamic range of 20 dB.}
		\label{3d_simo_mimo}
	\end{figure}
	
		\begin{figure}[!t]
		\centering
		\hfill    
		\subfloat[\label{simo_bpwk_ah}]{%
			\includegraphics[width=0.5\columnwidth]{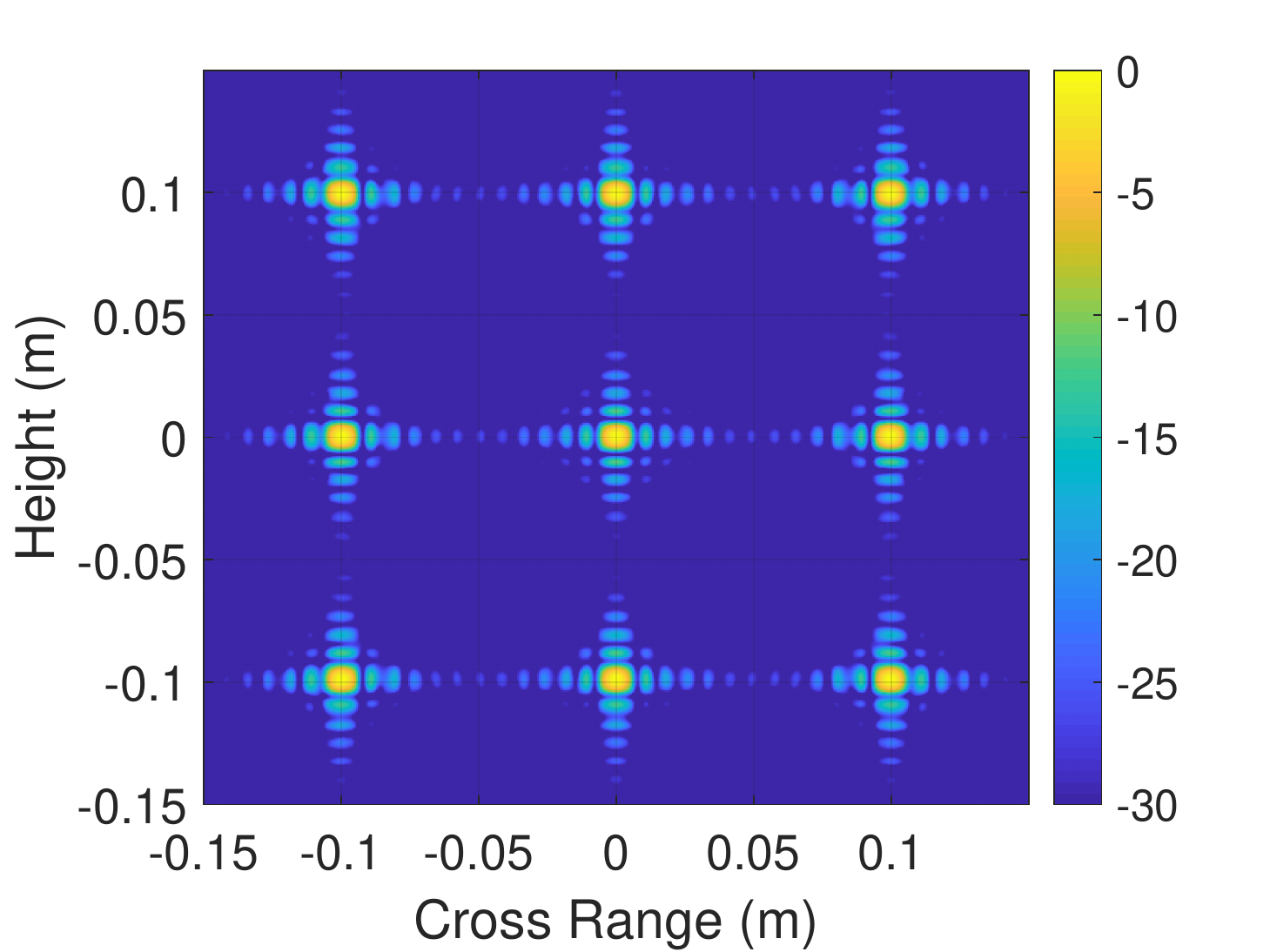}}
		\hfill
		\subfloat[\label{simo_bp_ah}]{%
			\includegraphics[width=0.5\columnwidth]{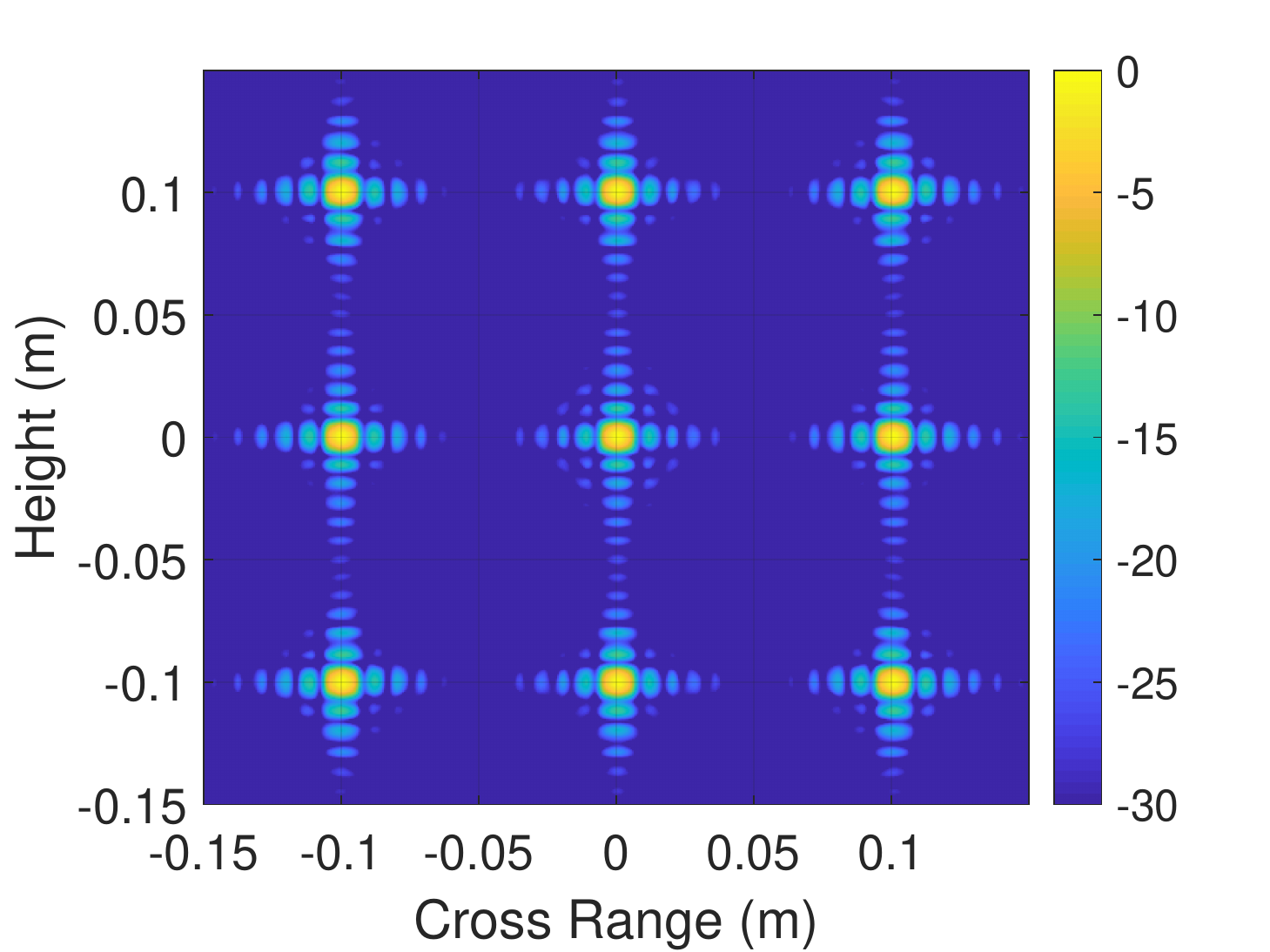}}
		\hfill
		    
		\subfloat[\label{simo_bpwk_ar}]{%
			\includegraphics[width=0.5\columnwidth]{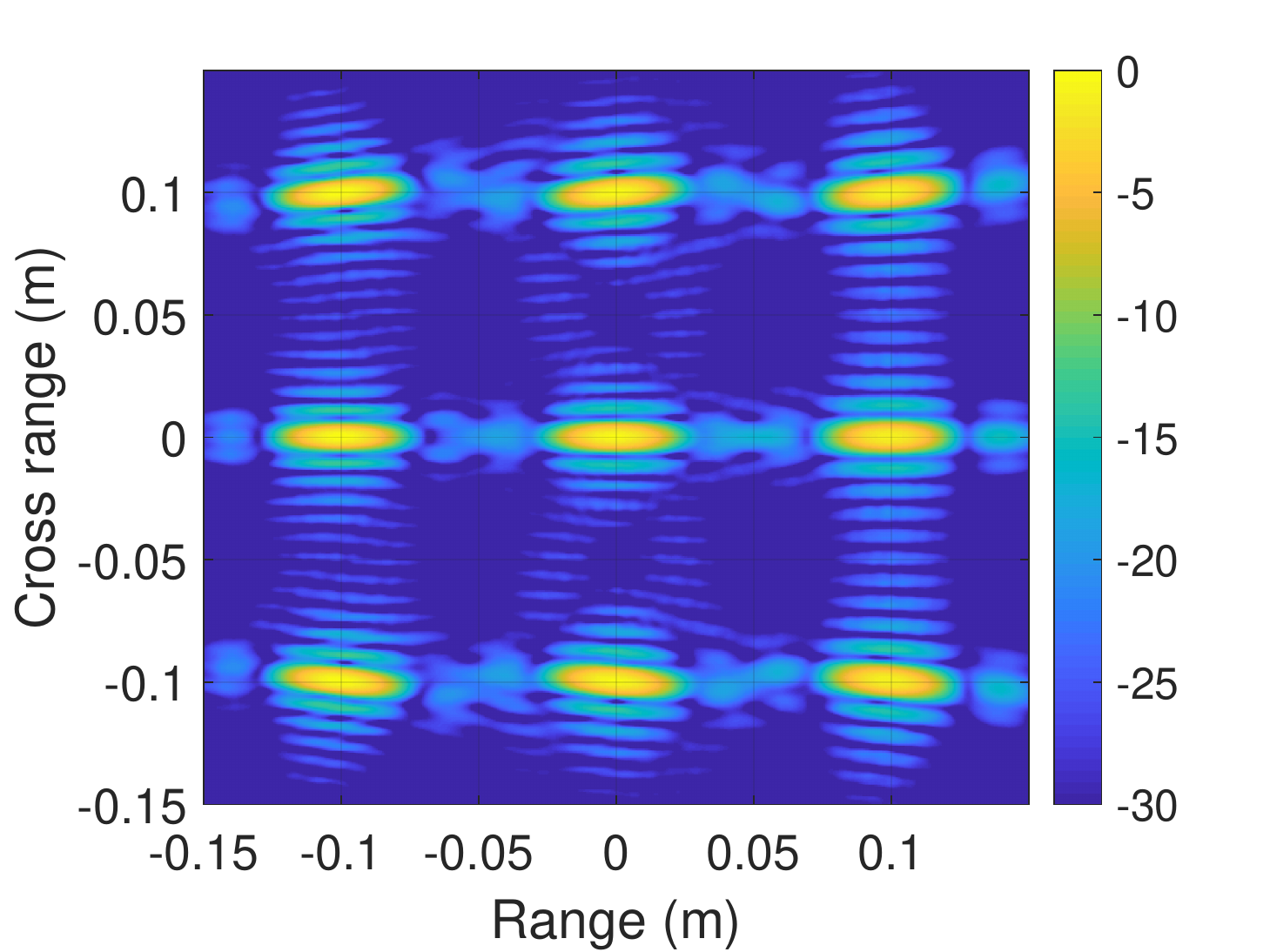}}
		\hfill
		\subfloat[\label{simo_bp_ar}]{%
			\includegraphics[width=0.5\columnwidth]{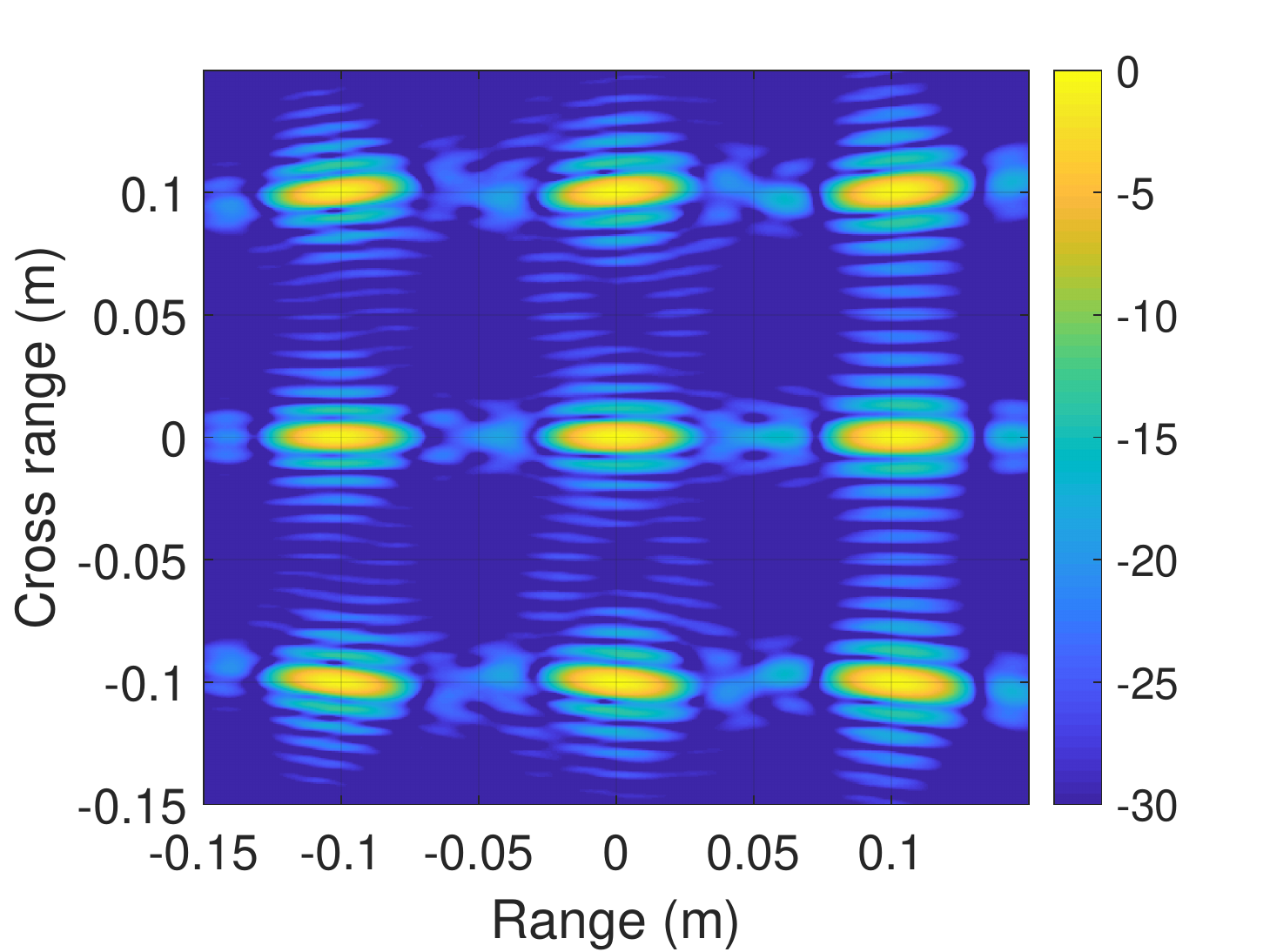}}

		\hfill    
		\subfloat[\label{simo_bpwk_rh}]{%
			\includegraphics[width=0.5\columnwidth]{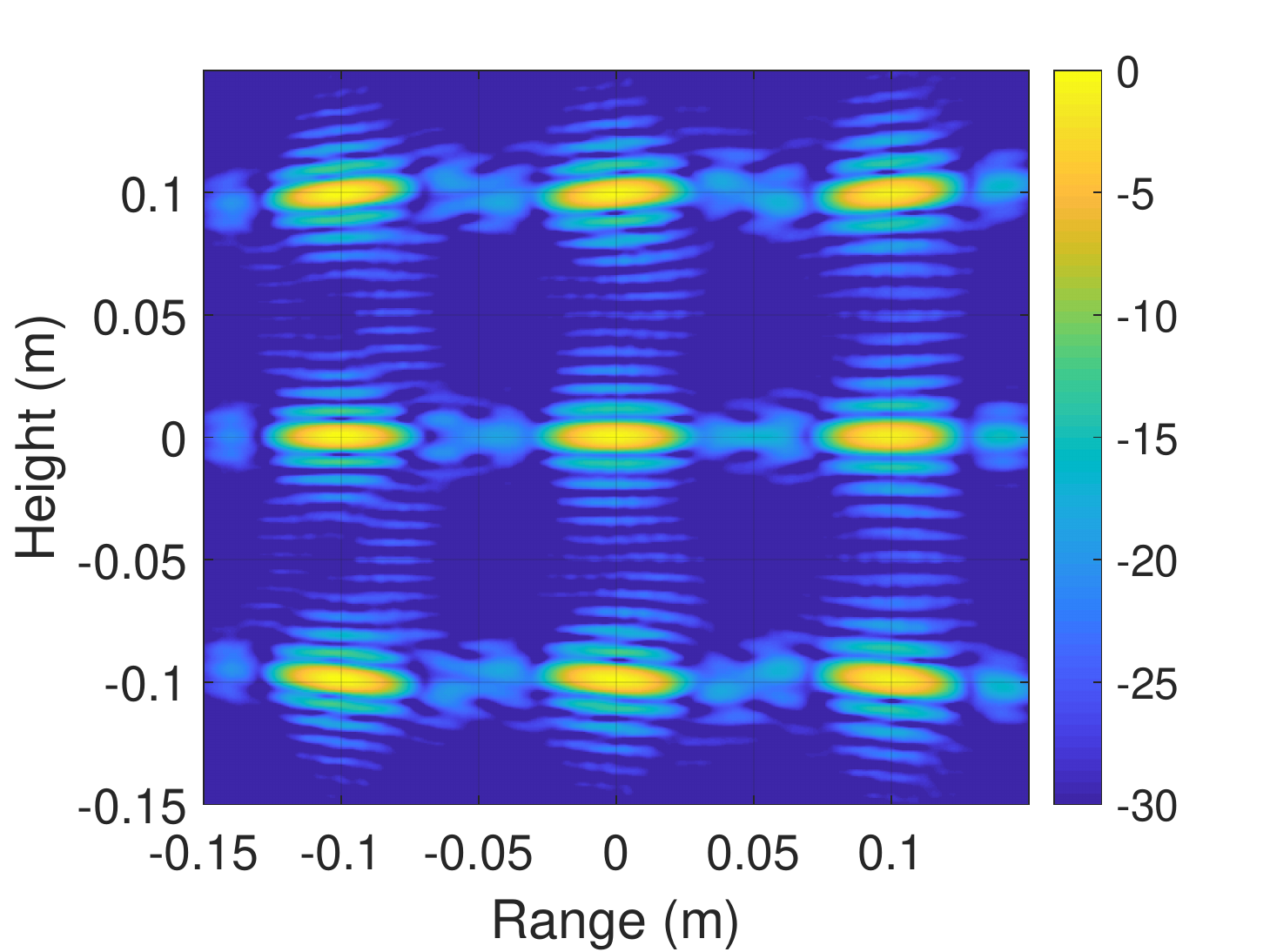}}
		\hfill
		\subfloat[\label{simo_bp_rh}]{%
			\includegraphics[width=0.5\columnwidth]{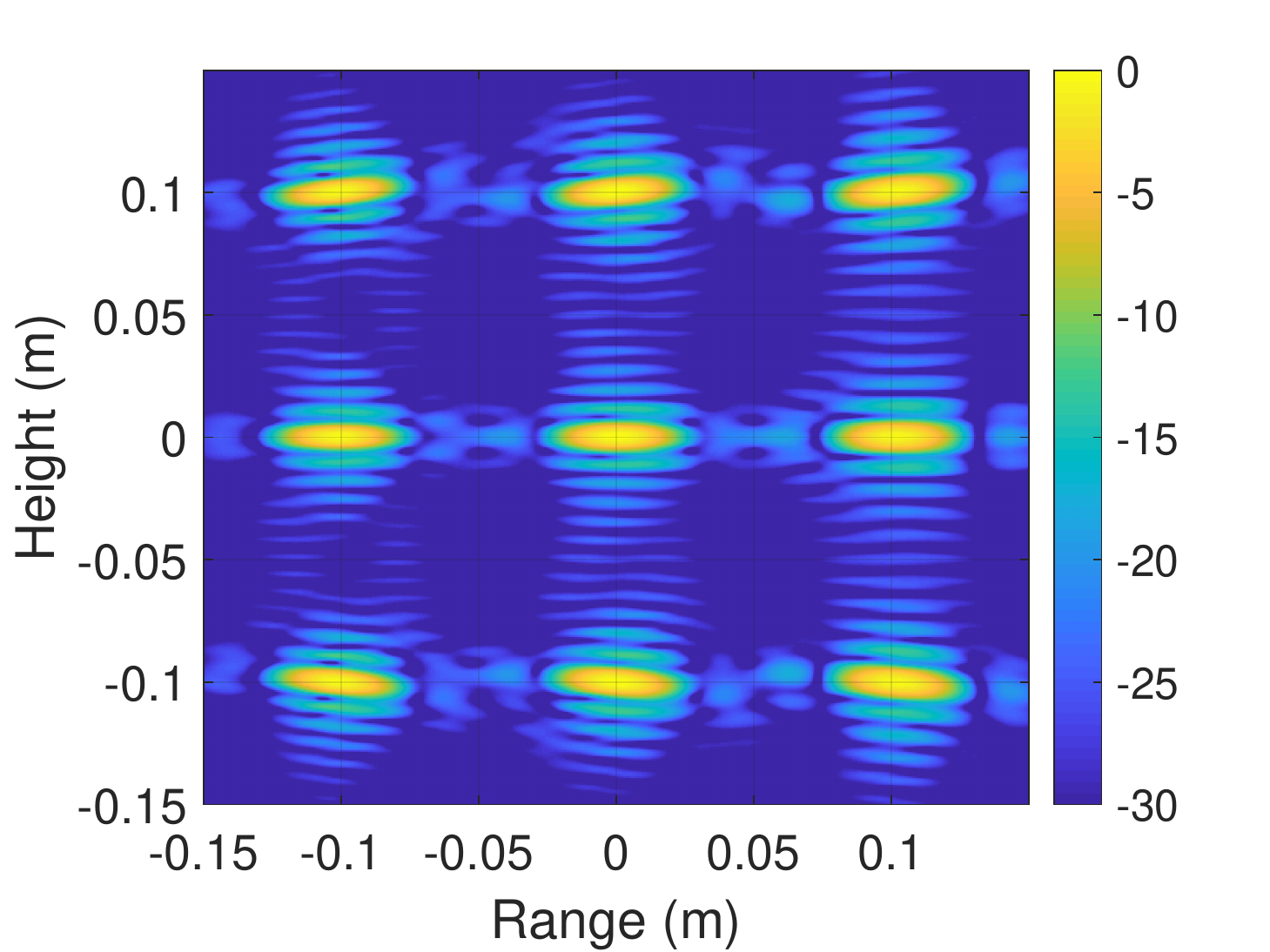}}
		\hfill

		\vspace{0mm}   
		
		\caption{2-D images with respect to the three coordinate planes by the proposed algorithm: (a), (c), (e), and by BP: (b), (d), (f),  via monostatic array.}
		\label{2-D slices_siso} 
	\end{figure}

		\begin{figure}[!t]
		\centering
	
		\hfill
		\subfloat[\label{mimo_bpwk_ah}]{%
			\includegraphics[width=0.5\columnwidth]{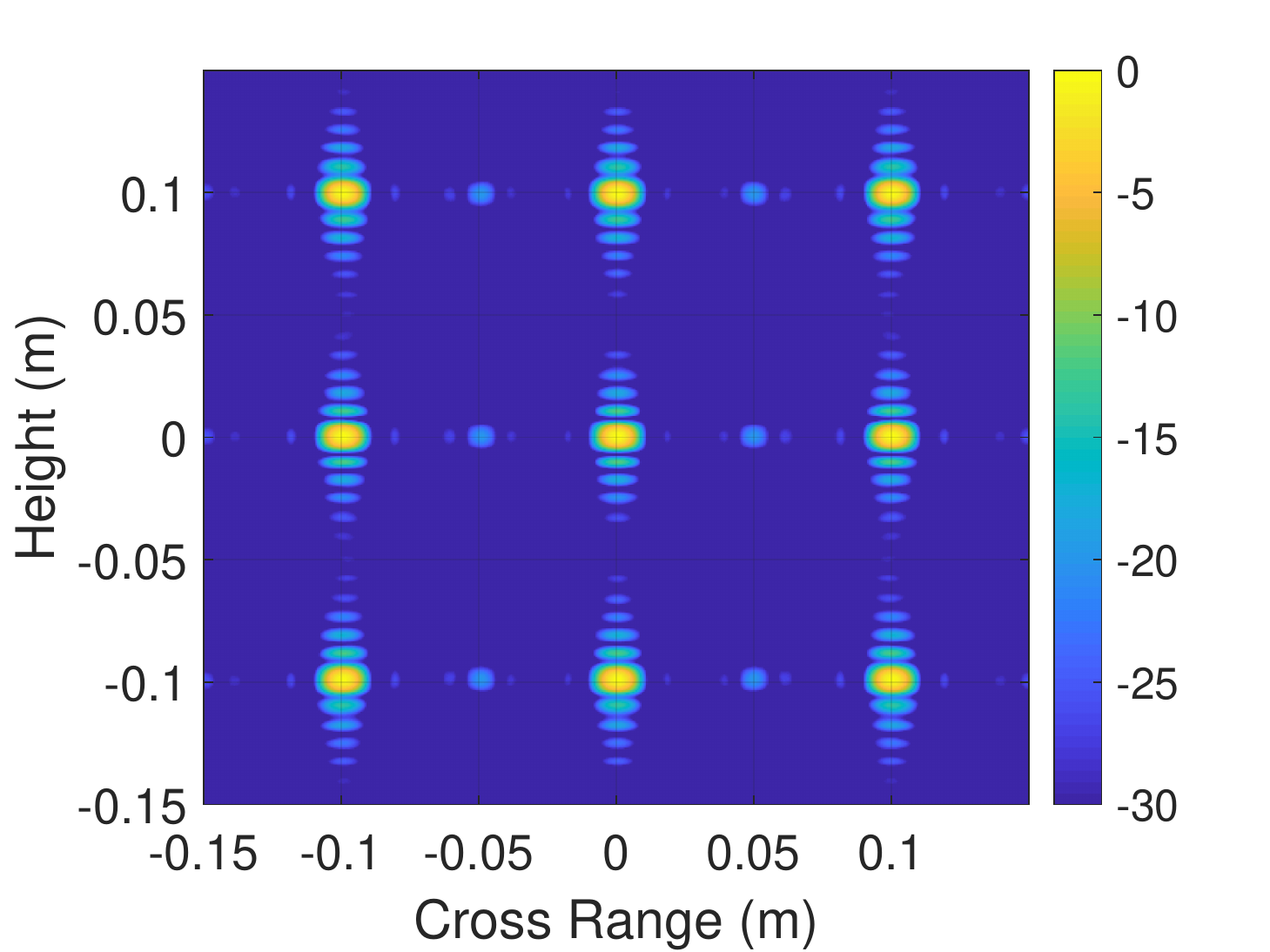}}
		\hfill
		\subfloat[\label{mimo_bp_ah}]{%
			\includegraphics[width=0.5\columnwidth]{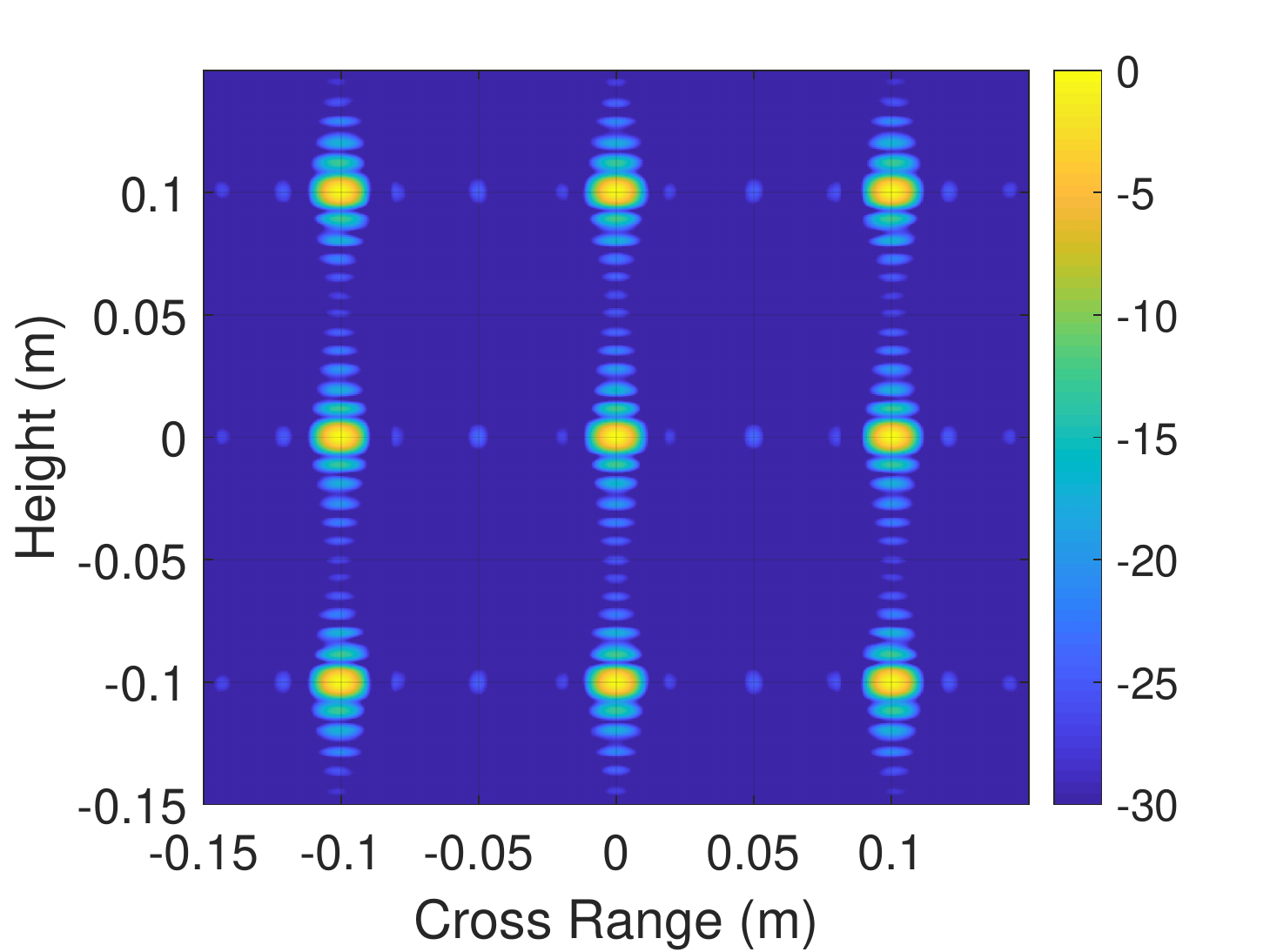}}     
		\hfill
	
		\subfloat[\label{mimo_bpwk_ar}]{%
			\includegraphics[width=0.5\columnwidth]{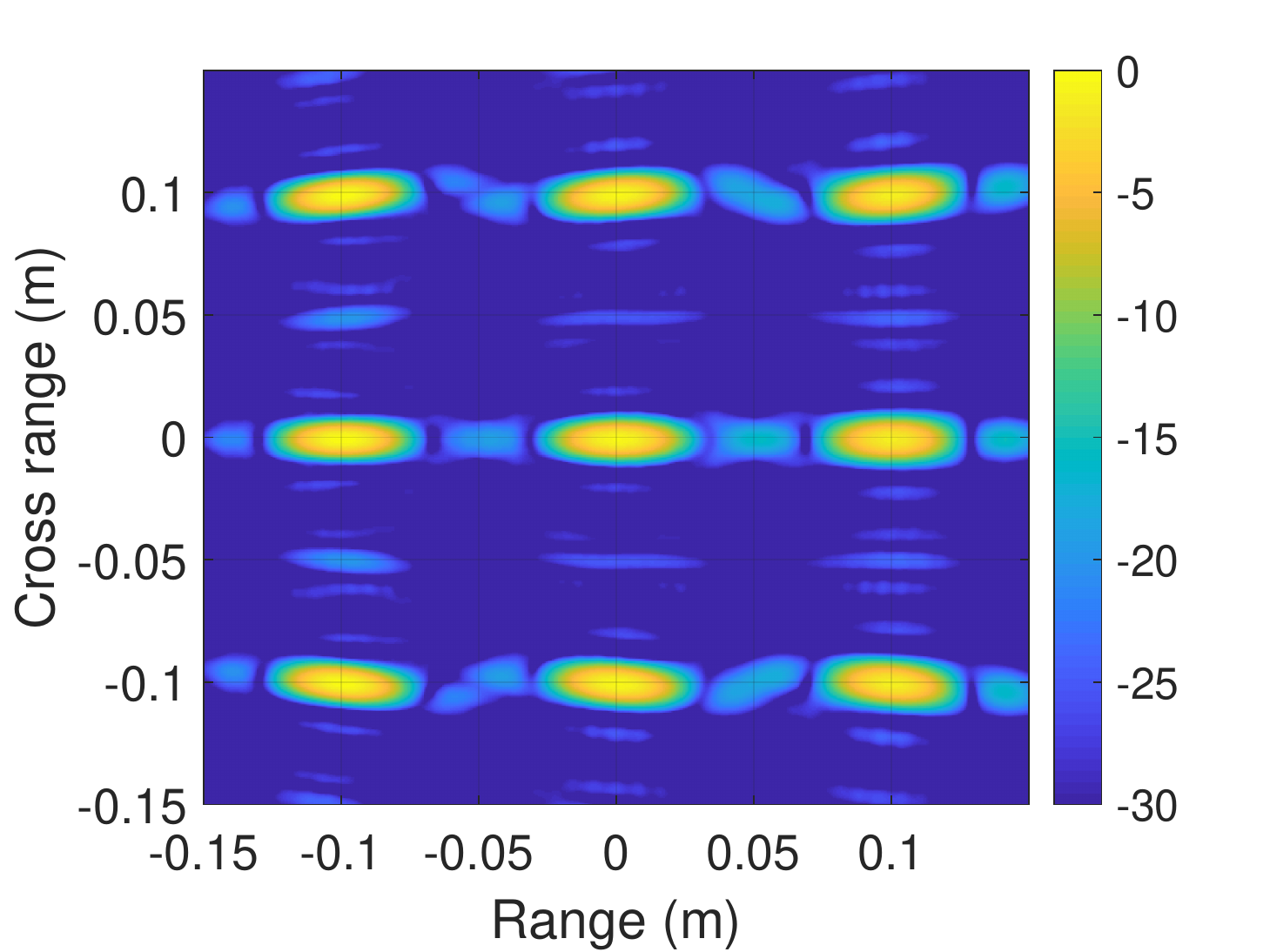}}
		\hfill
		\subfloat[\label{mimo_bp_ar}]{%
			\includegraphics[width=0.5\columnwidth]{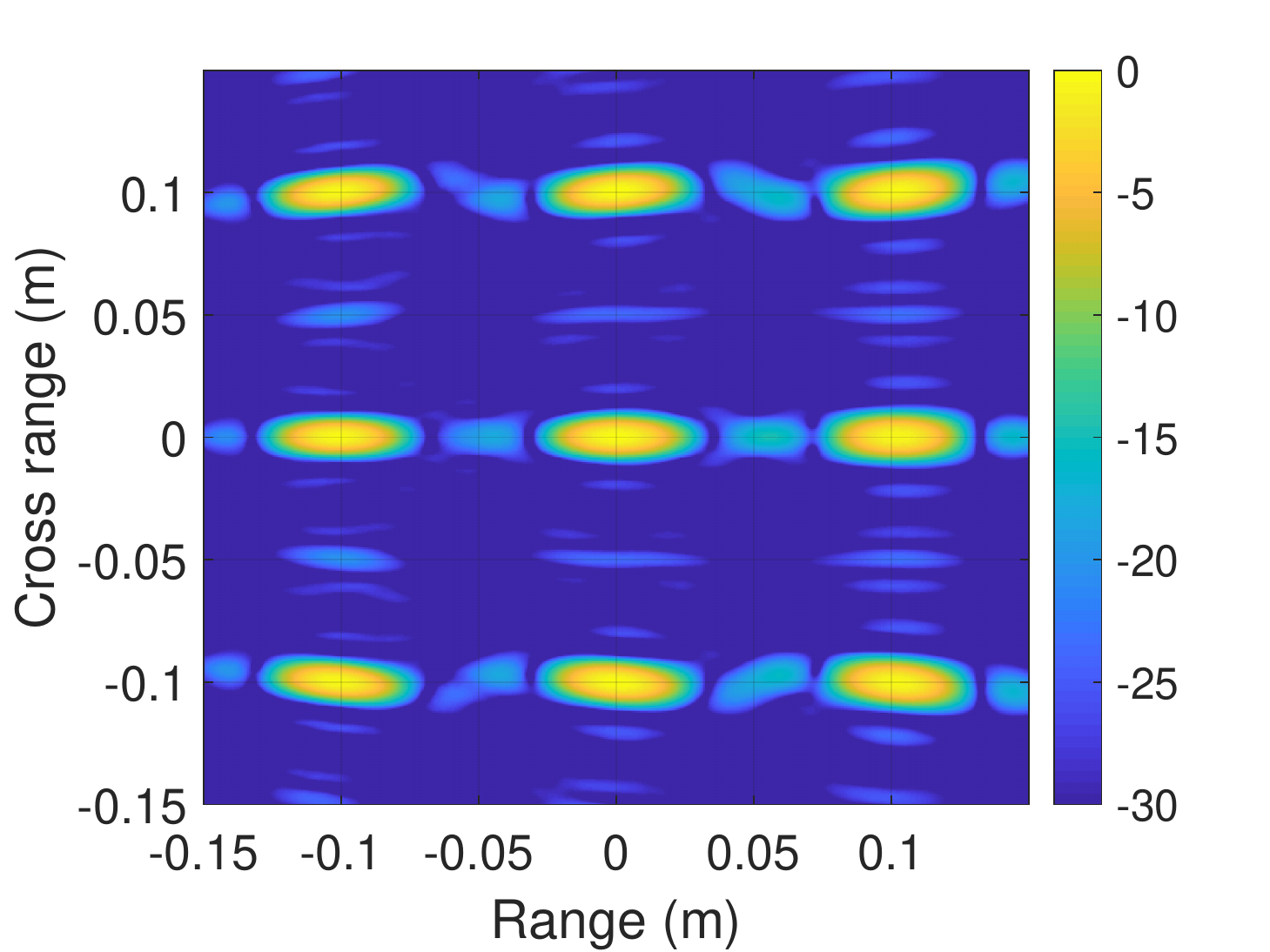}}     
		\hfill
	
		\subfloat[\label{mimo_bpwk_rh}]{%
			\includegraphics[width=0.5\columnwidth]{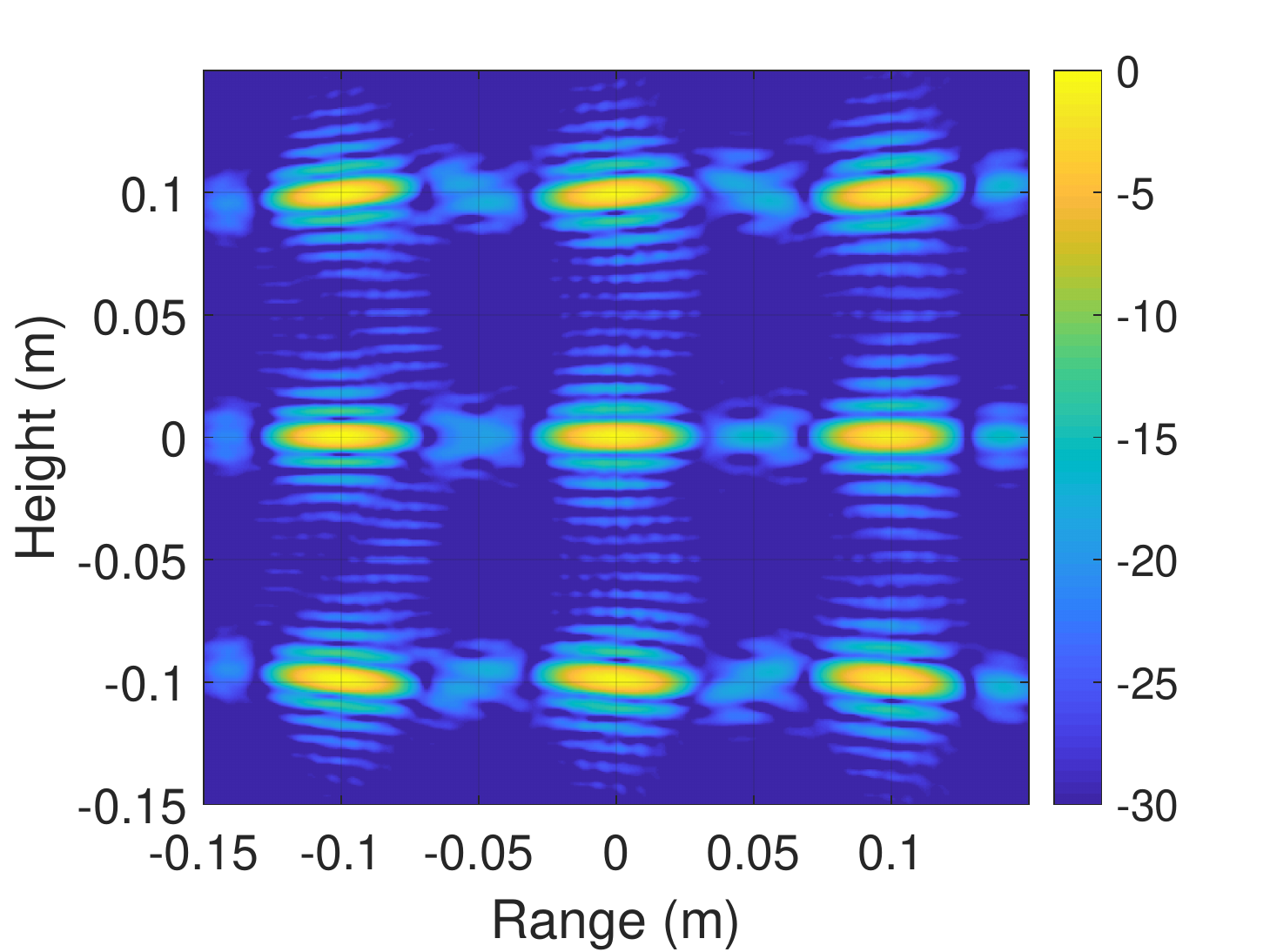}}
		\hfill
		\subfloat[\label{mimo_bp_rh}]{%
			\includegraphics[width=0.5\columnwidth]{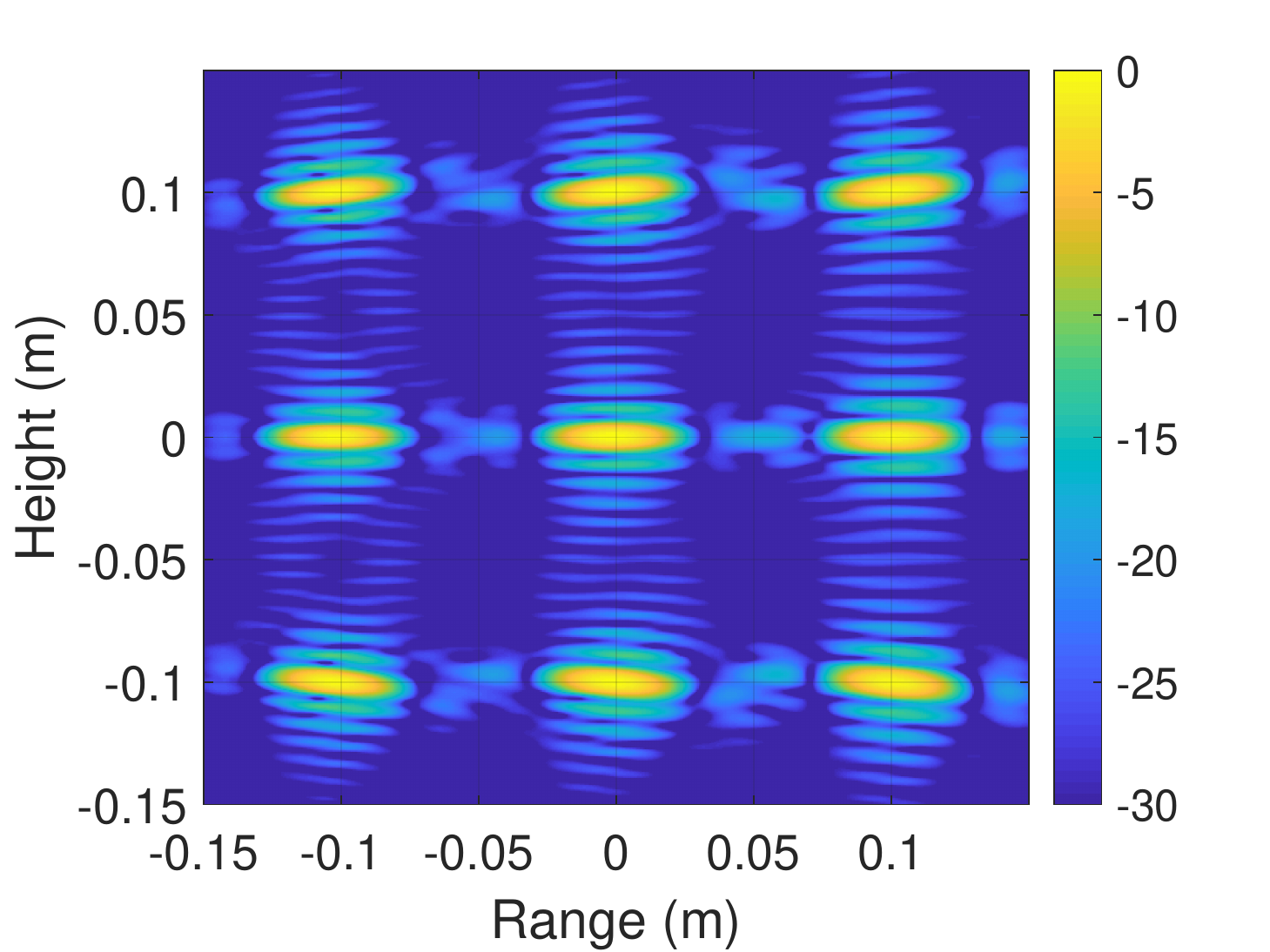}}     
		\hfill
		\vspace{0mm}   
		
		\caption{2-D images with respect to the three coordinate planes by the proposed algorithm: (a), (c), (e), and by BP: (b), (d), (f),  via multistatic array.}
		\label{2-D slices_mimo} 
	\end{figure}

		\begin{figure}[!t]
		\centering
		\subfloat[]{\label{a}
			\includegraphics[width=2.5in]{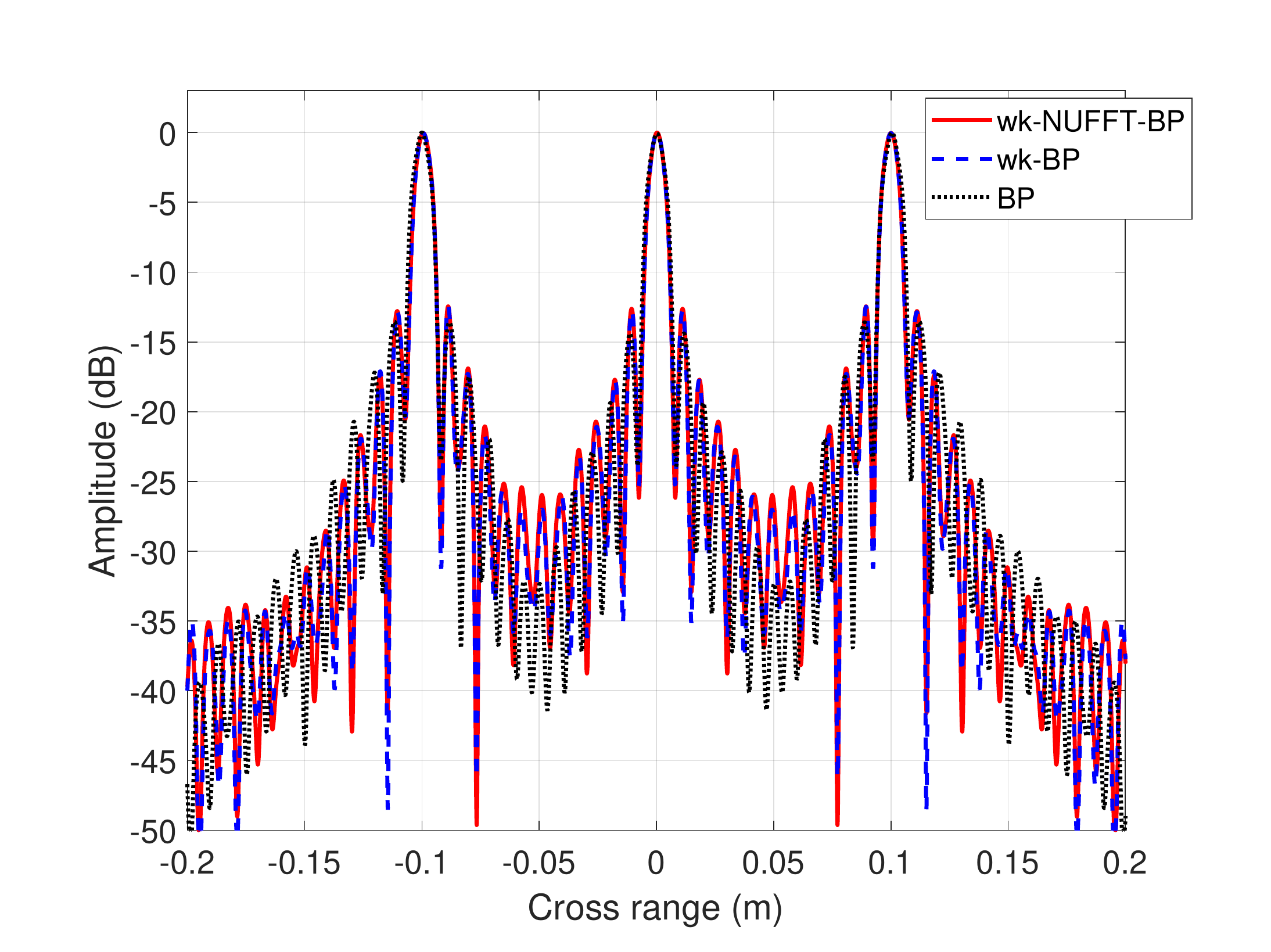}}
		\hfill
		\subfloat[]{\label{b}	
			\includegraphics[width=2.5in]{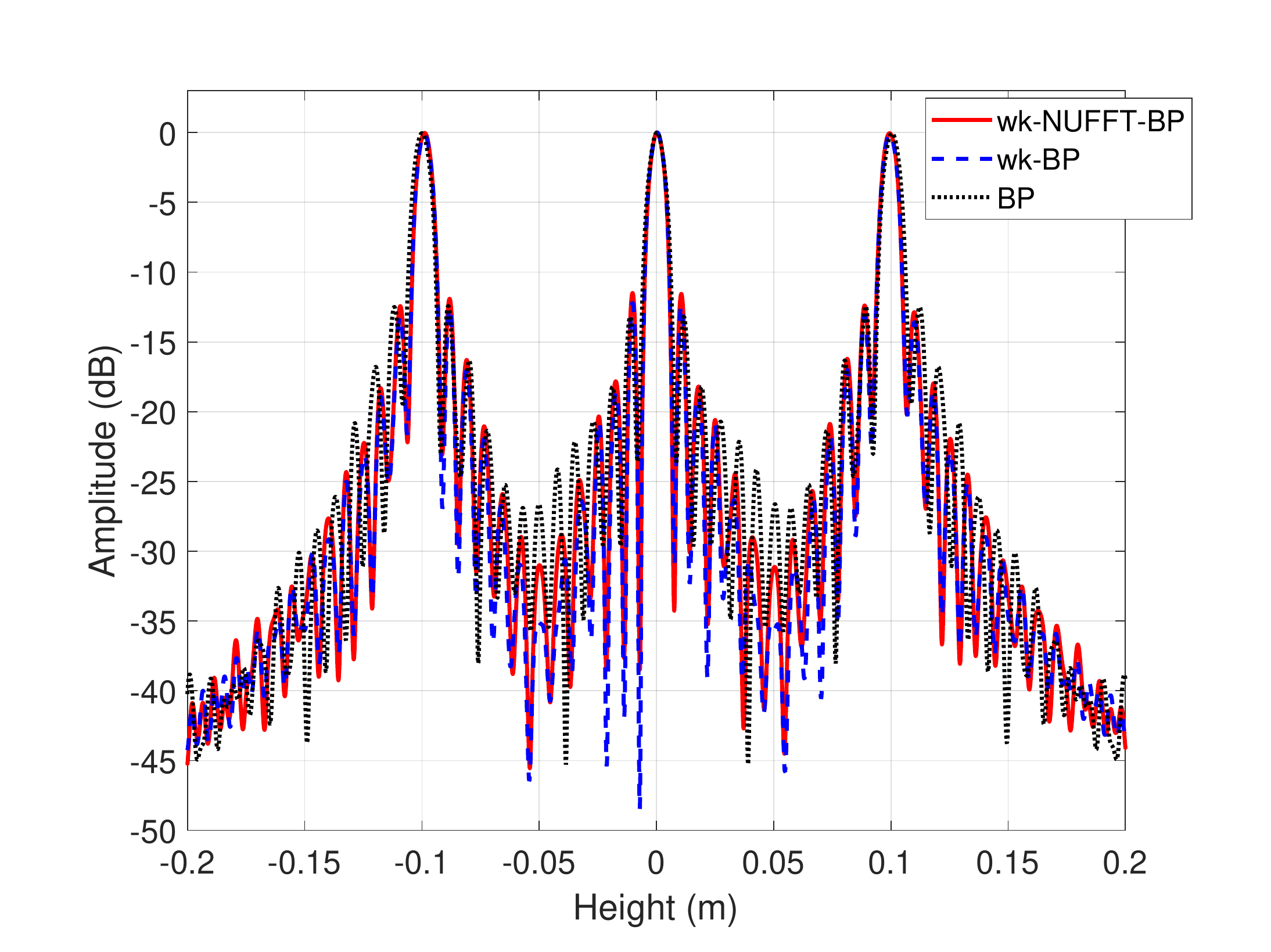}}
		\hfill
			\subfloat[]{\label{c}	
			\includegraphics[width=2.5in]{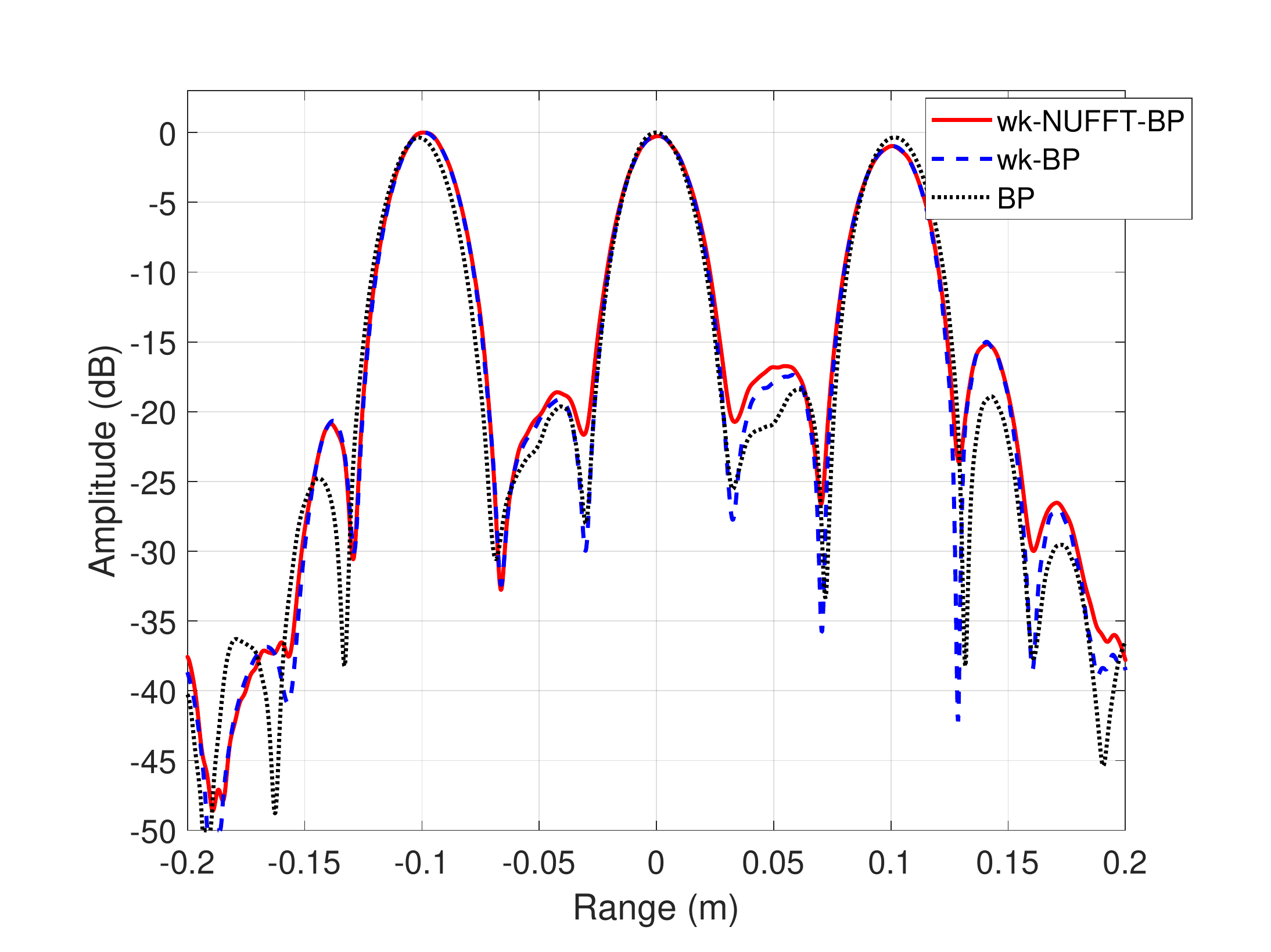}}	
		\hfill
		\\	
		\caption{1-D images with respect to the (a) horizontal, (b) height, and (c)
			range dimensions, via monostatic array.}
		\label{1-D-slices-mono}
	\end{figure}
	
	\begin{figure}[!t]
		\centering
		\subfloat[]{\label{a}
			\includegraphics[width=2.5in]{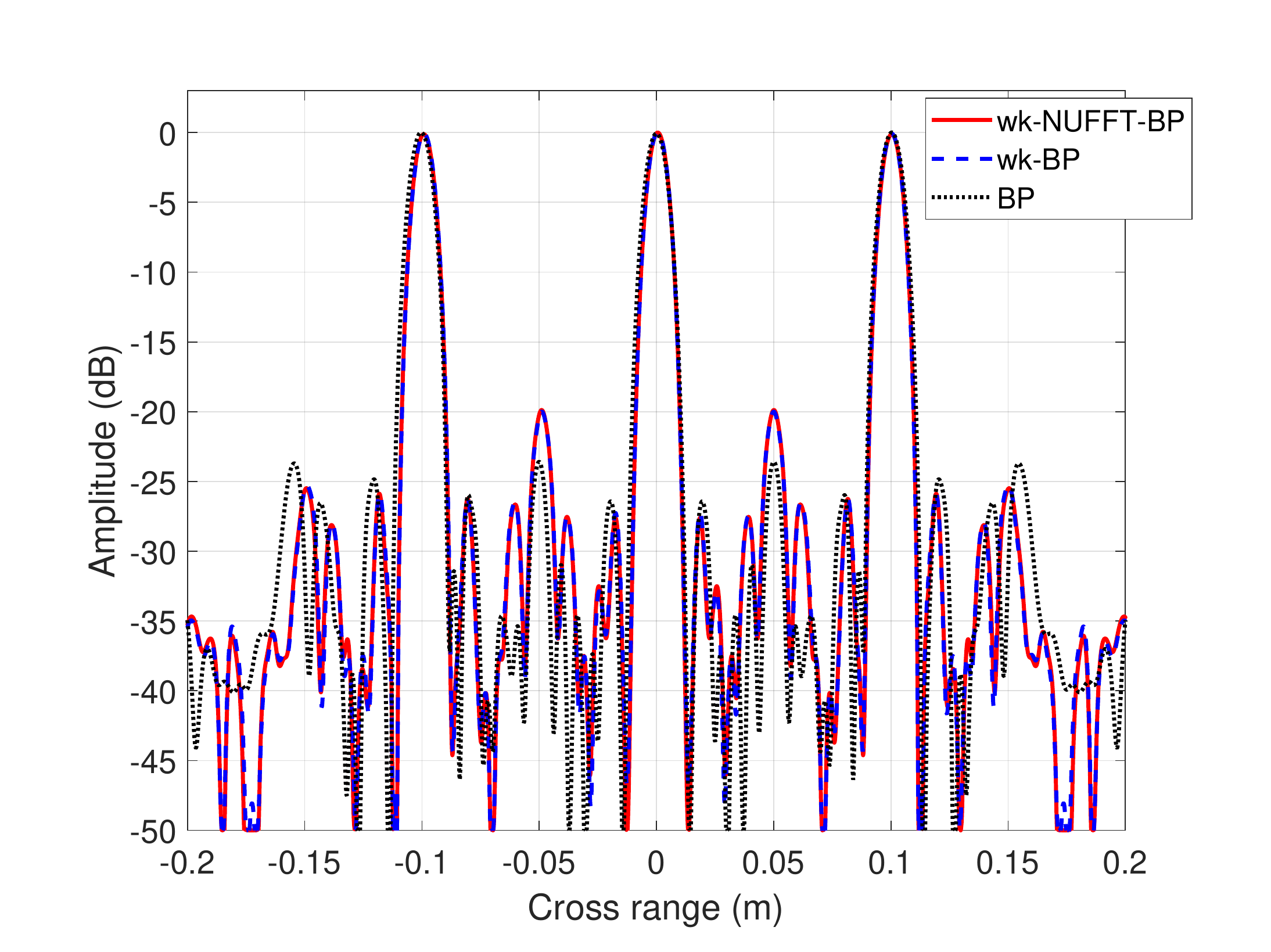}}
		\hfill
		\subfloat[]{\label{b}	
			\includegraphics[width=2.5in]{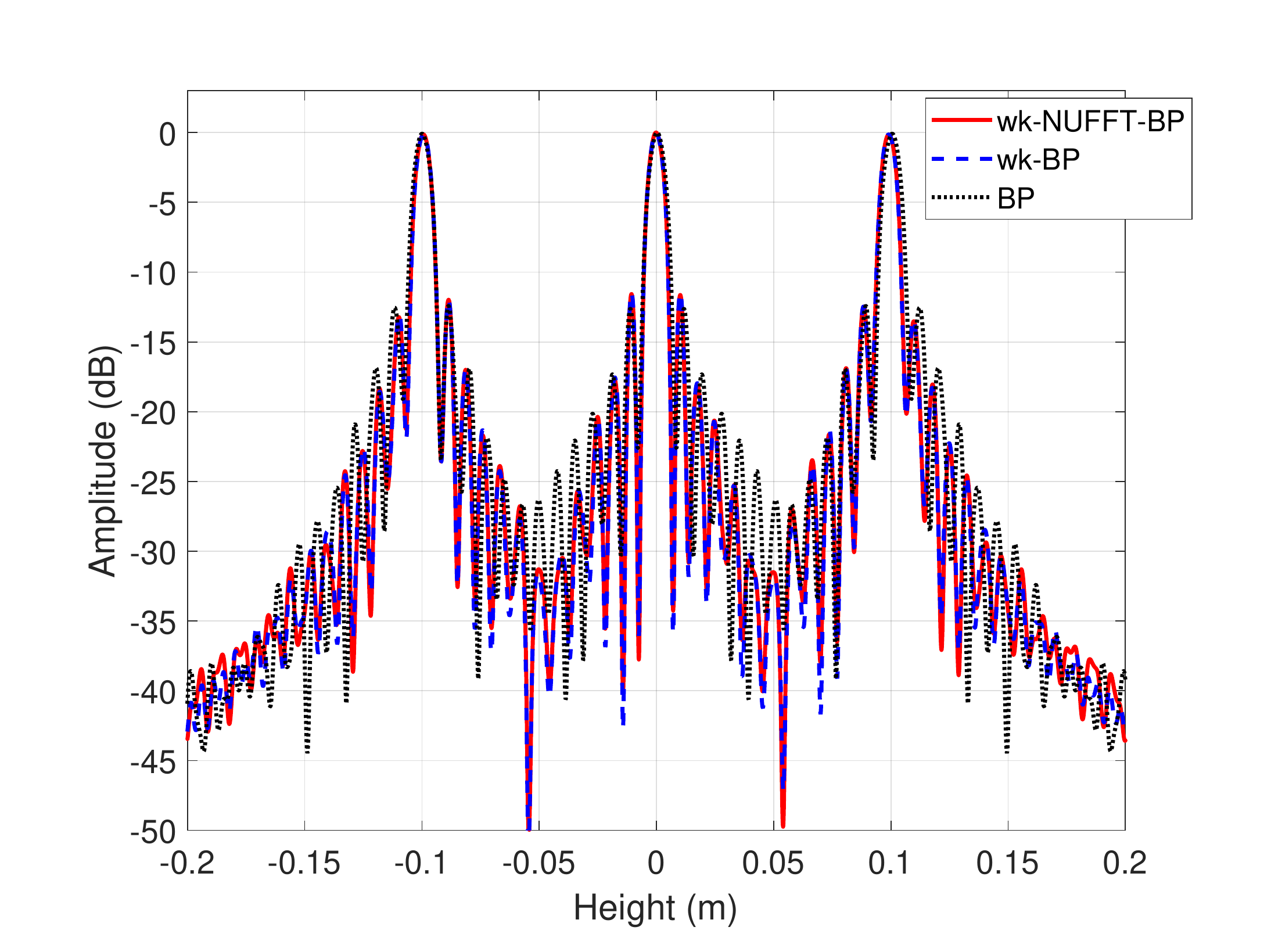}}	
		\hfill
		\subfloat[]{\label{c}	
			\includegraphics[width=2.5in]{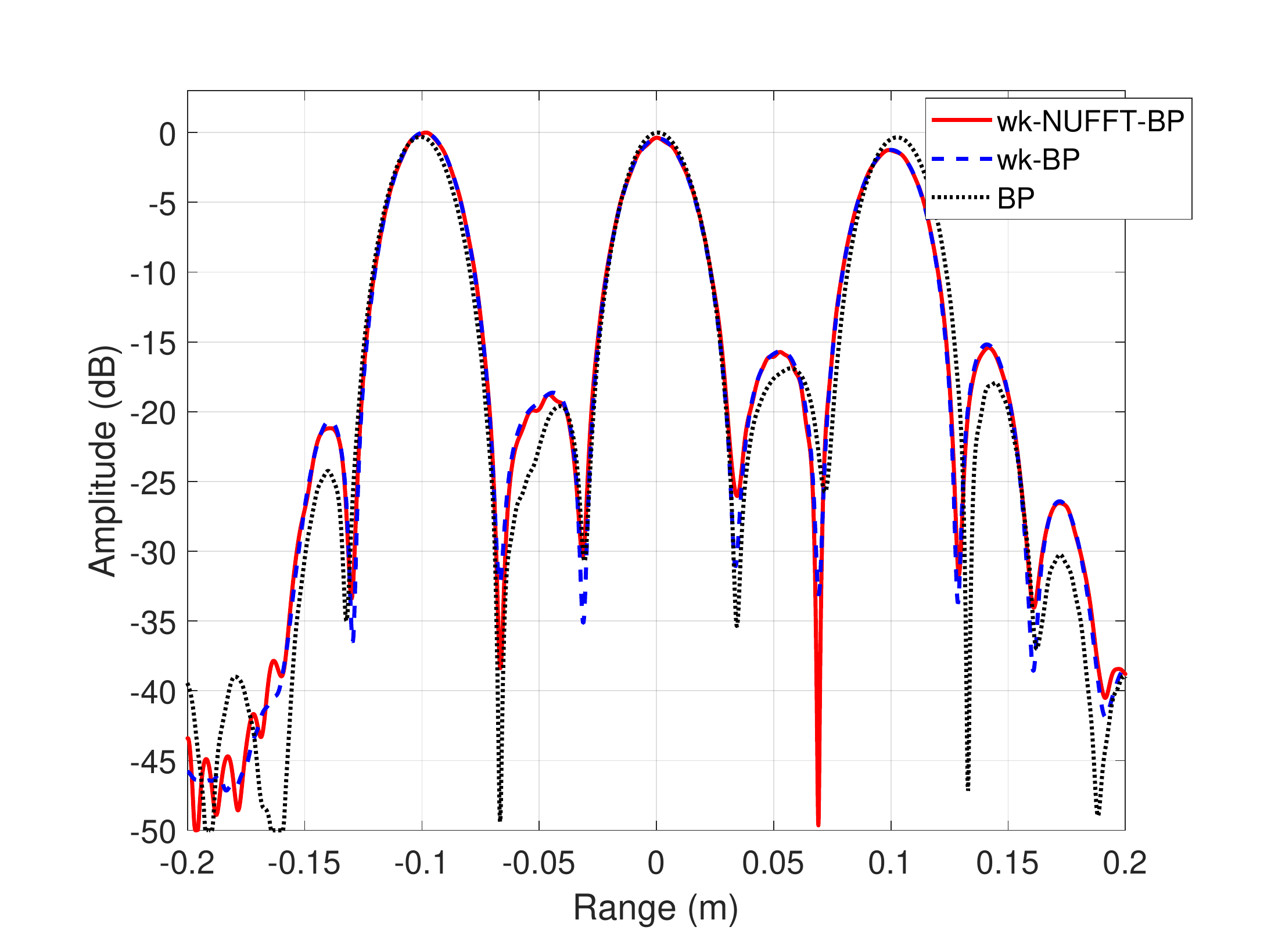}}	
		\hfill
		\\	
		\caption{1-D images with respect to the (a) horizontal, (b) height, and (c)
			range dimensions, via multistatic array.}
		\label{1-D-slices-mimo}
	\end{figure}

	We first provide imaging results of point targets to demonstrate the focusing performance in comparison with the BP algorithm. 
	Fig. \ref{3d_simo_mimo} shows the 3-D images using  the monostatic and multistatic arrays, respectively.
	To view the details, the corresponding 2-D and 1-D imaging results are shown in Figs. \ref{2-D slices_siso} to \ref{1-D-slices-mimo}, respectively. 
	In Figs. \ref{1-D-slices-mono} and \ref{1-D-slices-mimo}, we also illustrate the results without using NUFFT, denoted by `$\omega K$-BP'. The proposed algorithm is referred to as `$\omega K$-NUFFT-BP' for convenience.
	Clearly, the focusing property of the proposed algorithm is very similar to that of BP. Also, from the comparison between the monostatic and  multistatic results, we can see that the sidelobes of the latter along the horizontal cross-range direction are averagely lower than those of the former, as shown in Figs. \ref{1-D-slices-mono}\subref{a} and \ref{1-D-slices-mimo}\subref{a}, due to the multiplication of the beam patterns between the transmit array  and the receive array  for the multistatic system. 
	
	Table \ref{tab2} shows the computation time of the aforementioned three algorithms. 
	The algorithms were implemented in MATLAB using a workstation  with two pieces of 64-bit 2.90-GHz CPU (E5-2640v4). 
	Note that the proposed algorithm is much faster than both the algorithm without using NUFFT and the traditional BP. 
	
	\begin{table}
		\centering
		\caption{Compuation Time}
		\setlength{\tabcolsep}{3pt}
		\begin{threeparttable}
			\begin{tabular}{p{70pt}  p{60pt} p{30pt}  p{30pt}}
				\hline\hline
				Algorithms& $\omega K$-NUFFT-BP& $\omega K$-BP& BP \\[0.5ex]
				\hline		
				time (monostatic)& 28 s& 370 s& 2779 s \\[0.5ex]
				time (multistatic)&  62 s& 685 s& 5824 s \\[0.5ex]
				\hline
			\end{tabular}
		\end{threeparttable}
		\label{tab2}
	\end{table}

	\begin{figure}[!t]
		\centering
		\includegraphics[width=2in]{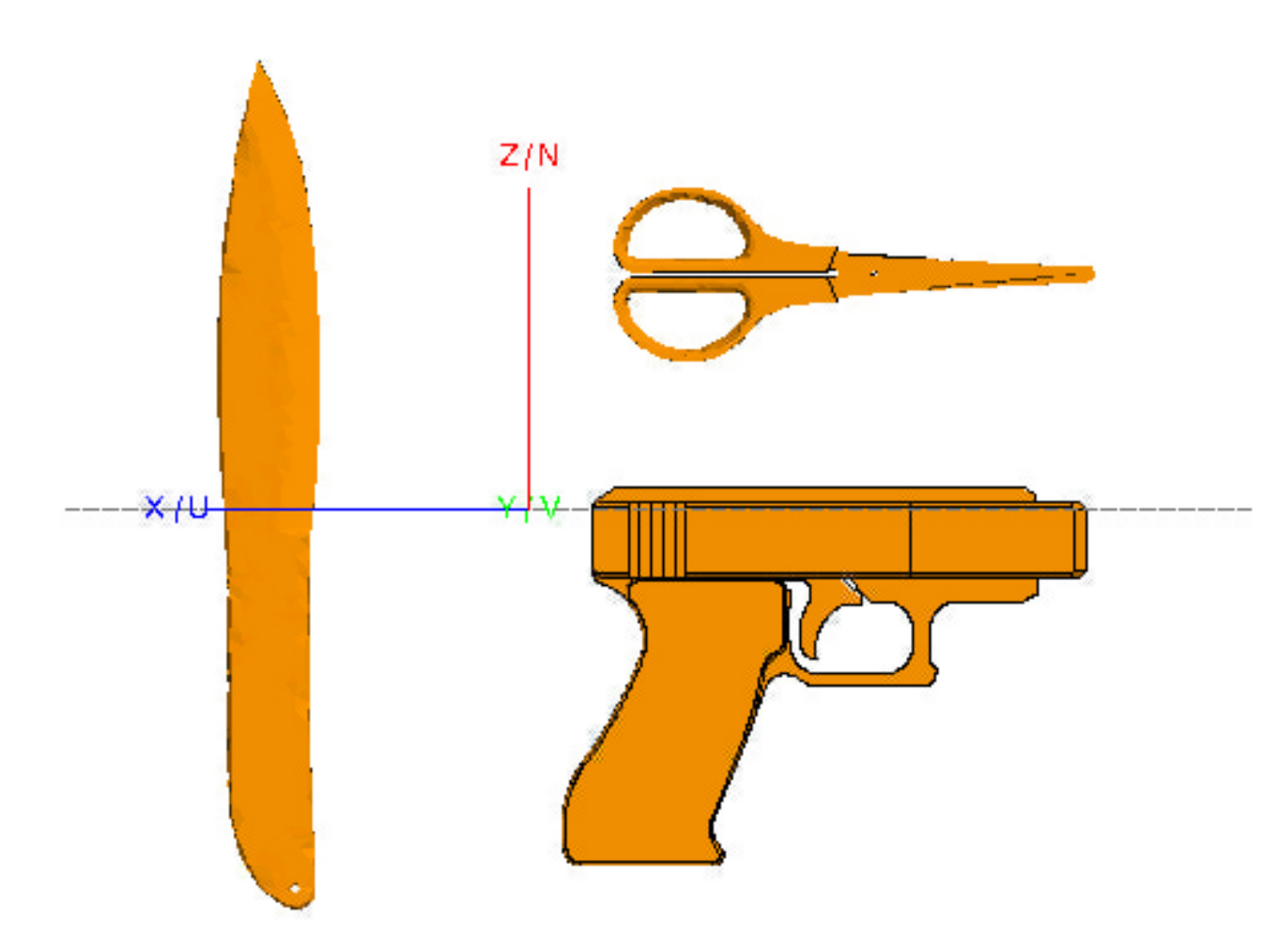}
		
		\caption{Target model in Feko.}
		\label{feko_ori}
	\end{figure}
	
	\begin{figure}[!t]
		\centering
		\subfloat[]{\label{a}
			\includegraphics[width=1.69in]{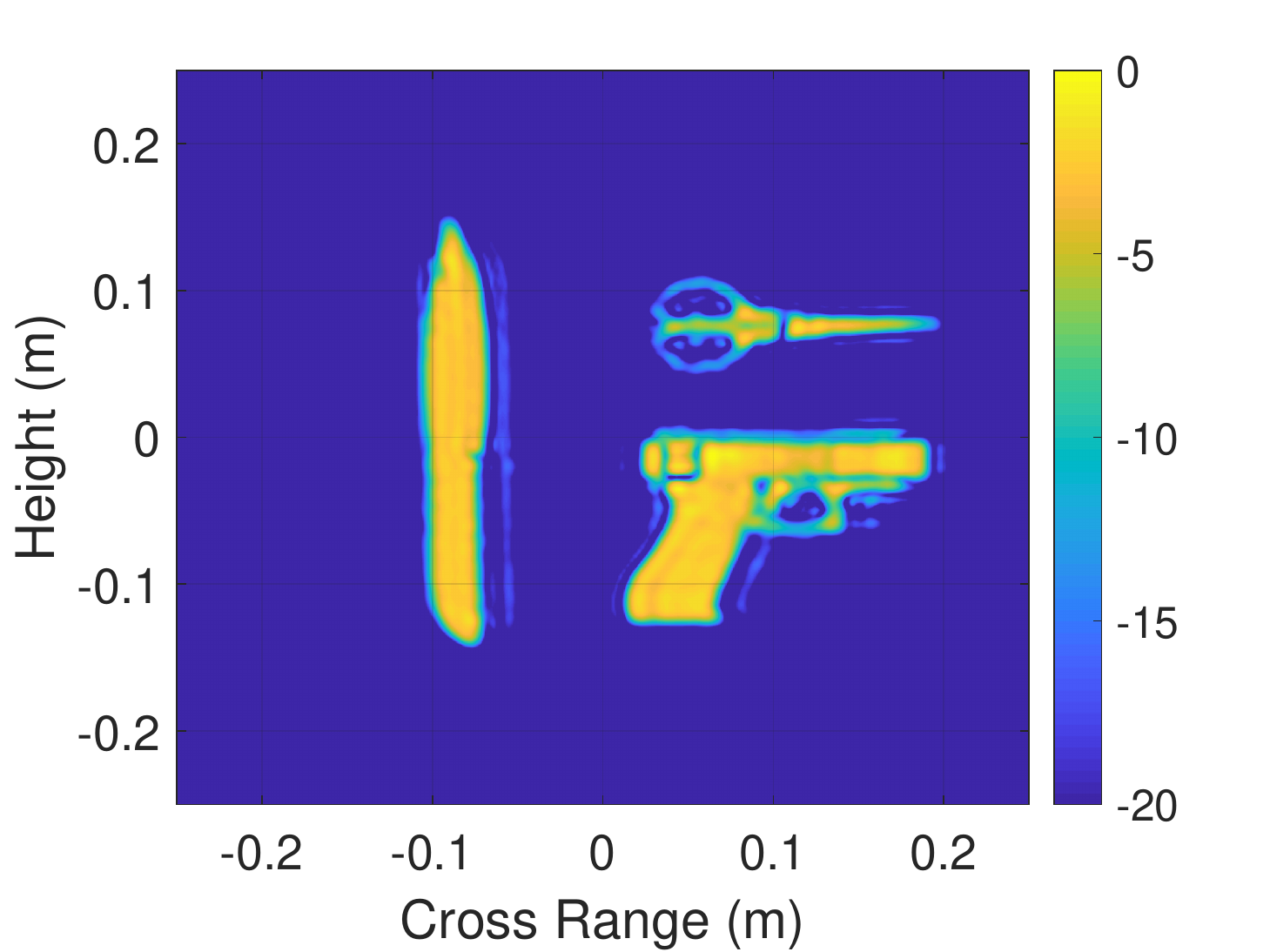}}
		\hfill
		\subfloat[]{\label{b}
			\includegraphics[width=1.69in]{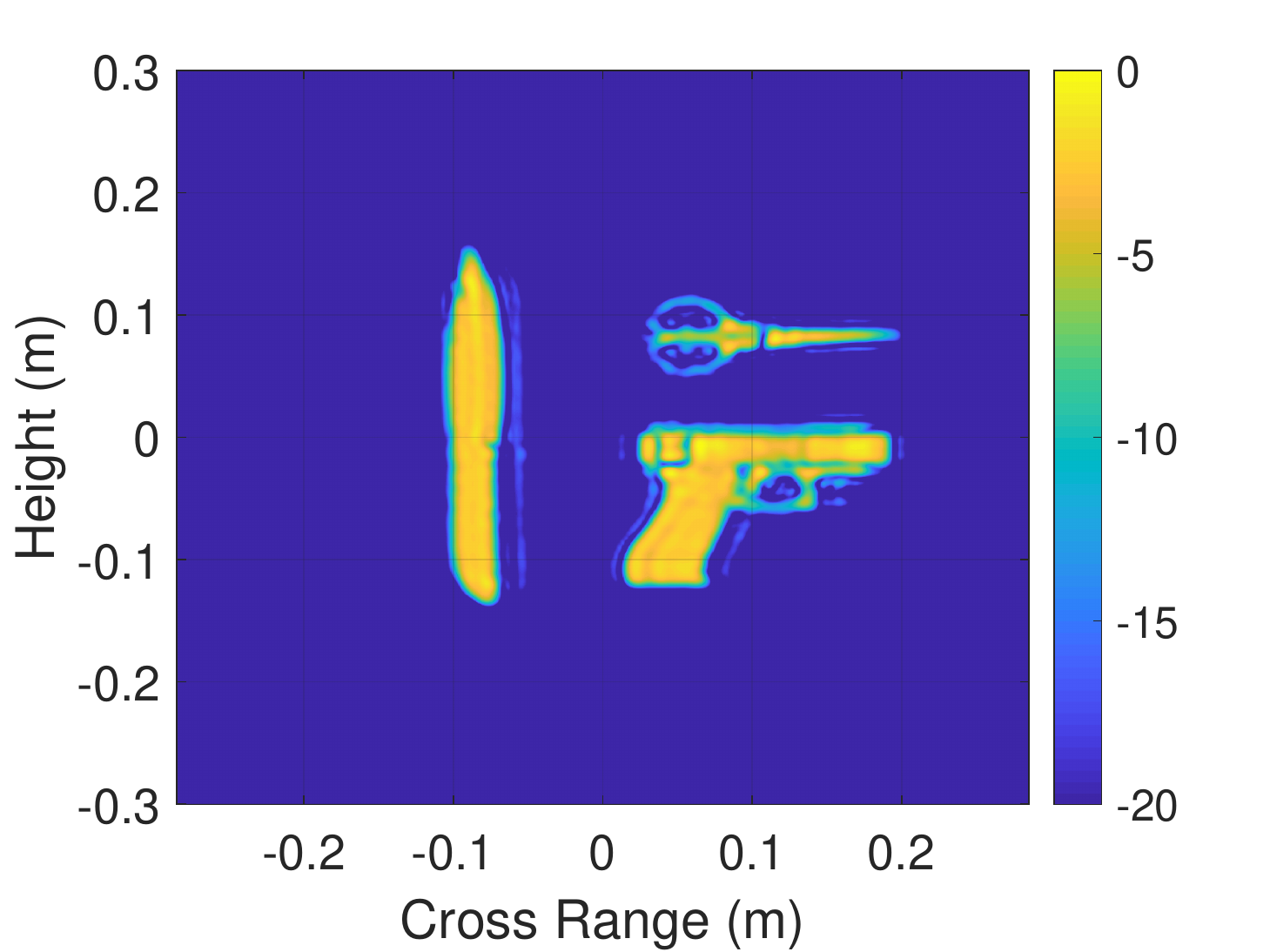}}
		\hfill
		\centering
		\subfloat[]{\label{c}
			\includegraphics[width=1.69in]{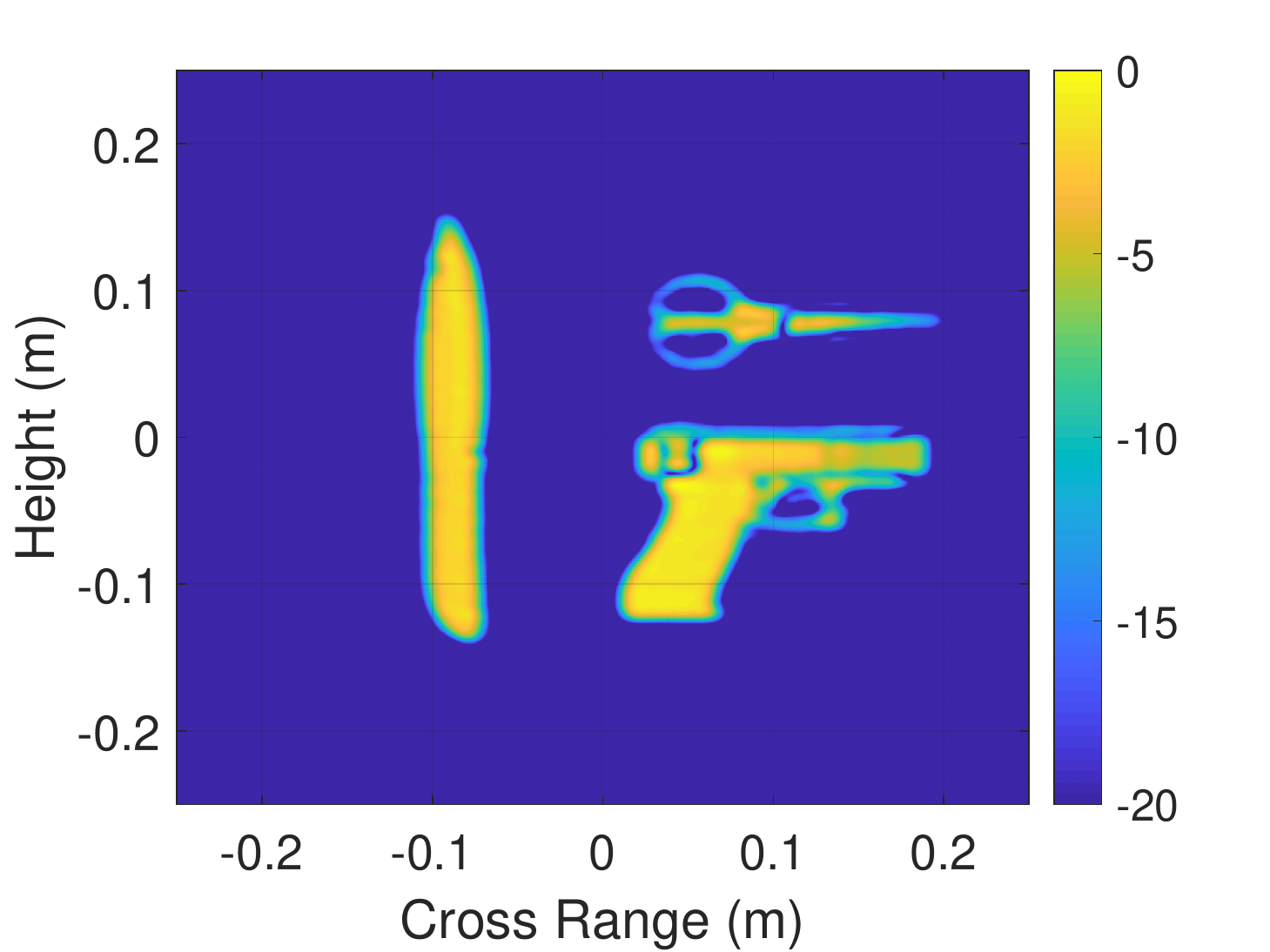}}
		\hfill
		\subfloat[]{\label{d}
			\includegraphics[width=1.69in]{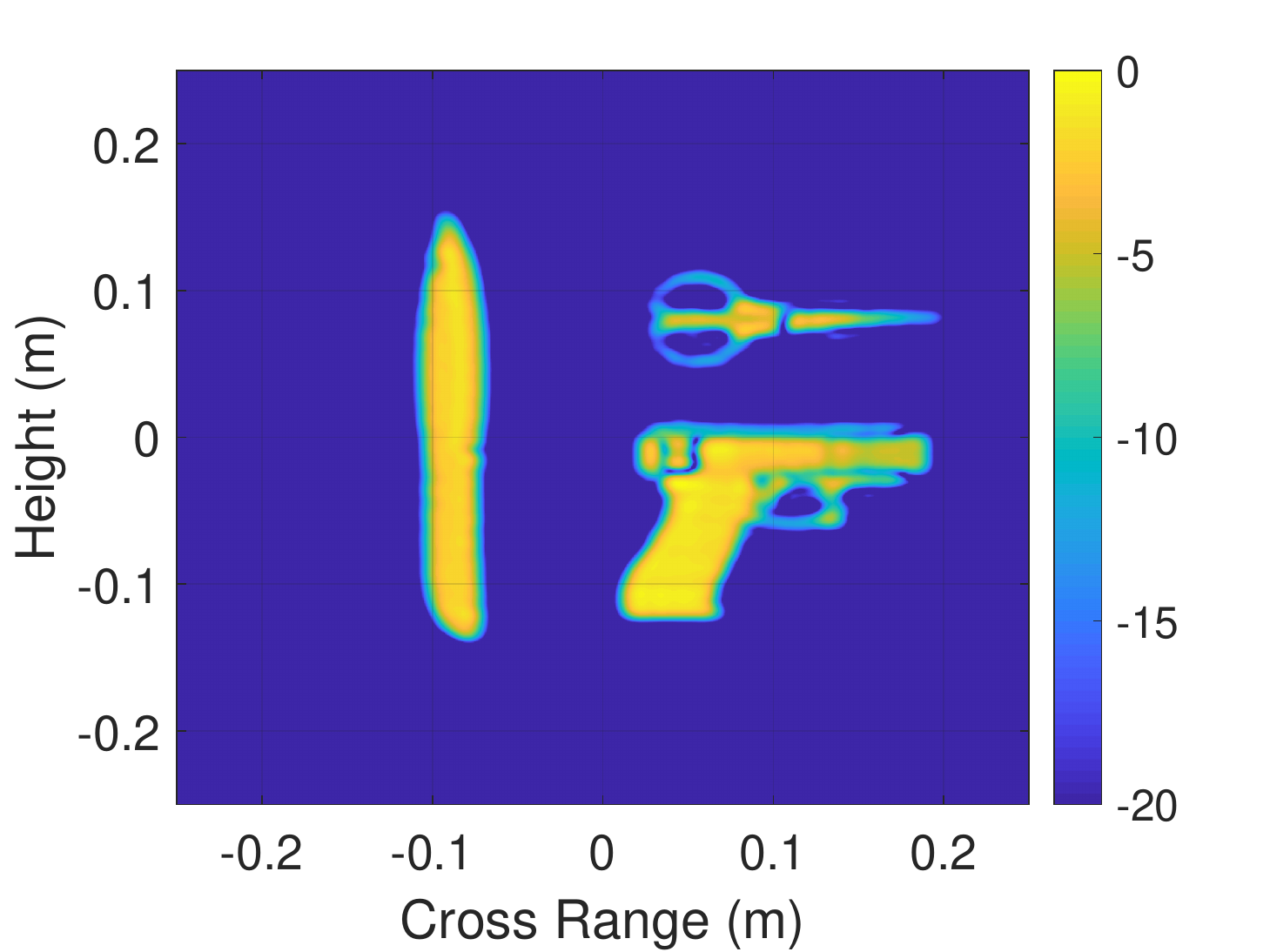}}
		\hfill
		\\	
		\caption{Imaging results  via  monostatic array by (a) the proposed algorithm, and (b) BP; and via multistatic array by (c) the proposed algorithm, and (d) BP.}
		\label{feko}
	\end{figure}
	
	Further, we employ FEKO - a computational electromagnetics software \cite{feko}, to simulate the scattered EM waves.
	The target model is illustrated in Fig. \ref{feko_ori}. The reconstructed images  are shown in Fig. \ref{feko}, by using the proposed algorithm and BP, respectively, for both the monostatic and multistatic array-based systems. 
	It is further shown that the proposed algorithm exhibits similar performance with that of BP, however, with much faster computational speed.

	\begin{figure}[!t]
	\centering
	\subfloat[]{\label{a}
		\includegraphics[width=1.69in]{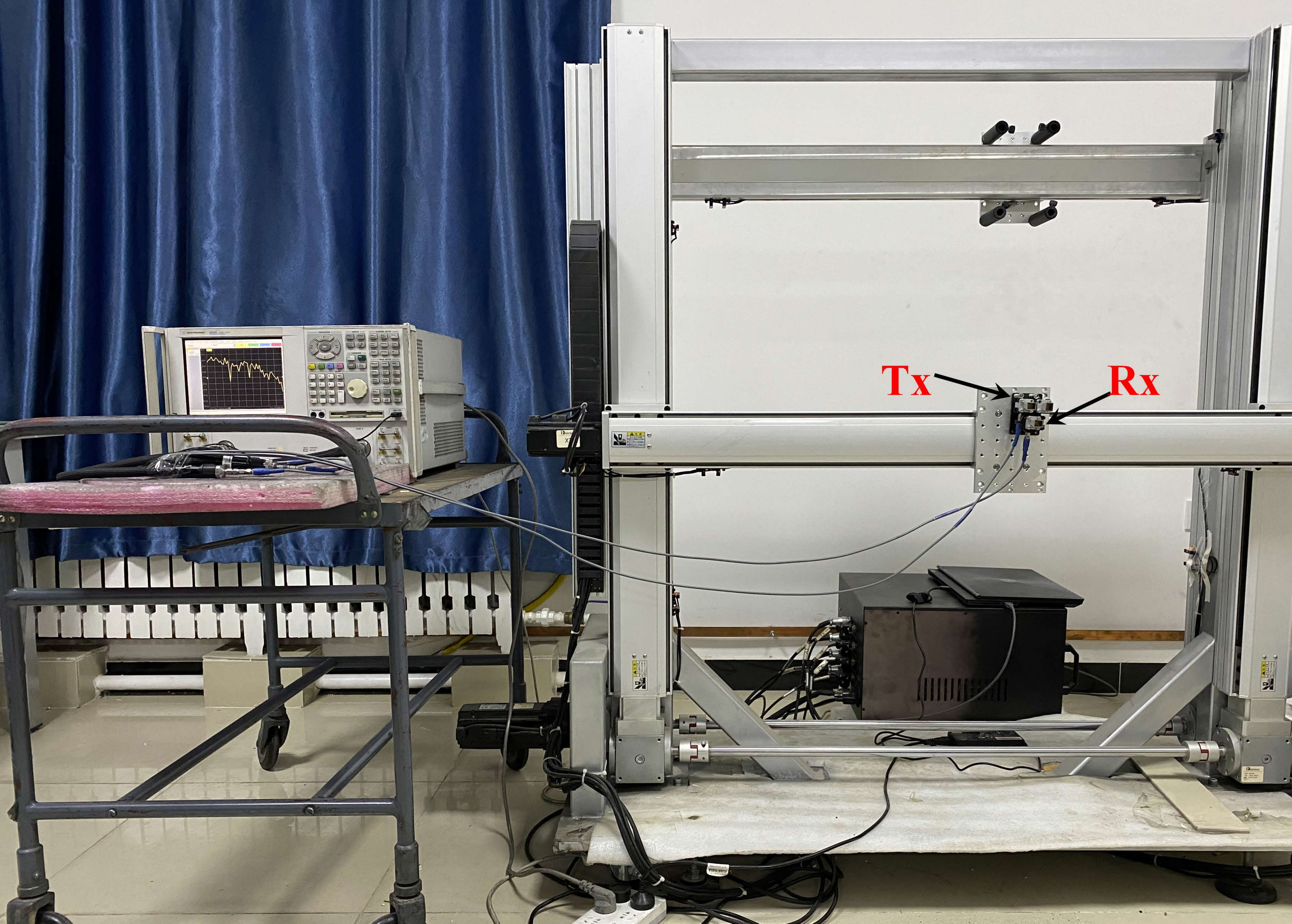}}
	\hfill
	\subfloat[]{\label{b}
		\includegraphics[width=1.69in]{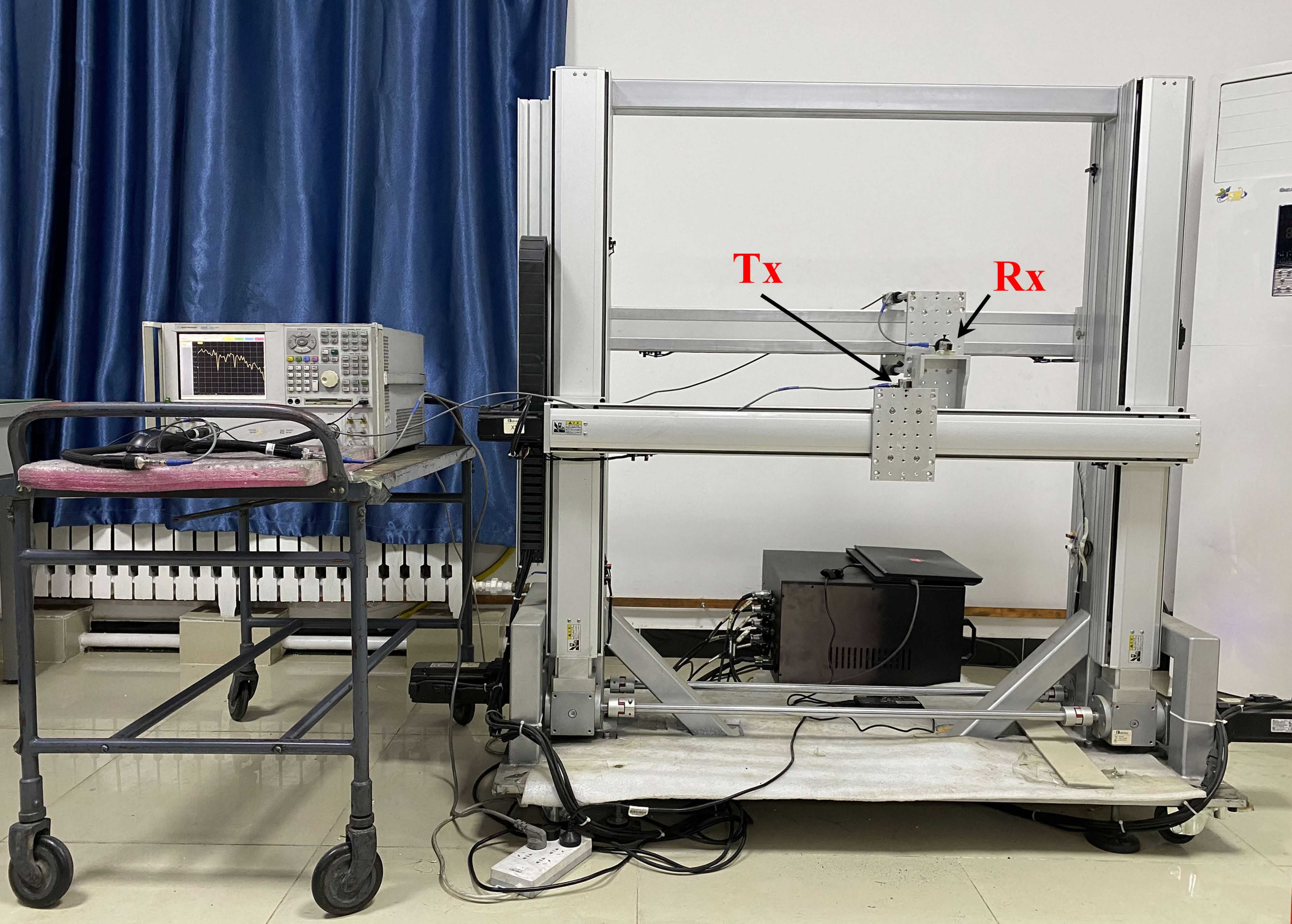}}
	\hfill
	\centering
	\subfloat[]{\label{c}
		\includegraphics[width=1.69in]{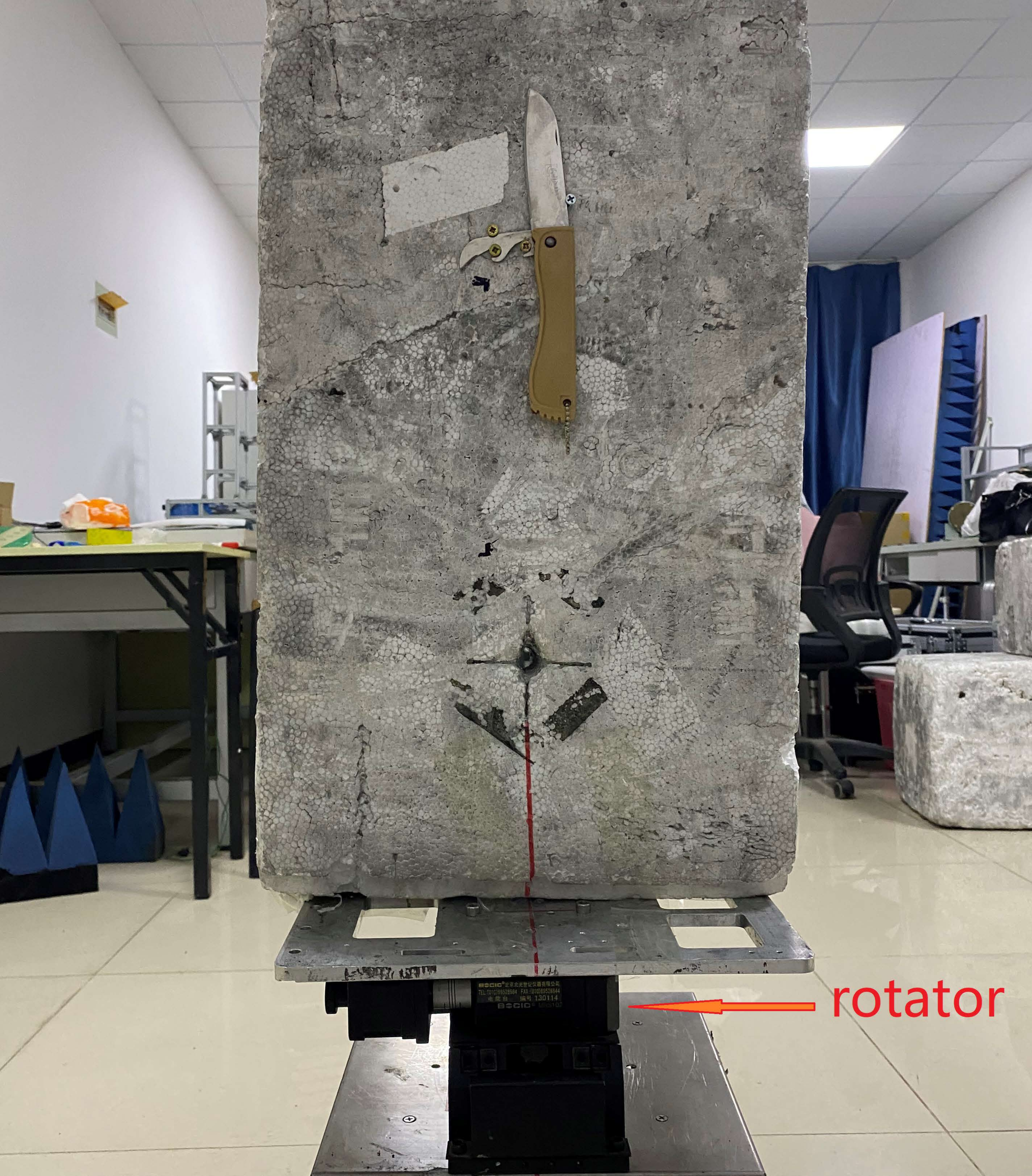}}

	\caption{Experiments setup, (a) the monostatic scenario, (b) the multistatic scenario, and (c) the target (a knife) fixed to a rotator.}
	\label{experiment_setup}
	
    \end{figure}
		\begin{figure}[!t]
		\centering
		\subfloat[]{\label{a}
			\includegraphics[width=1.69in]{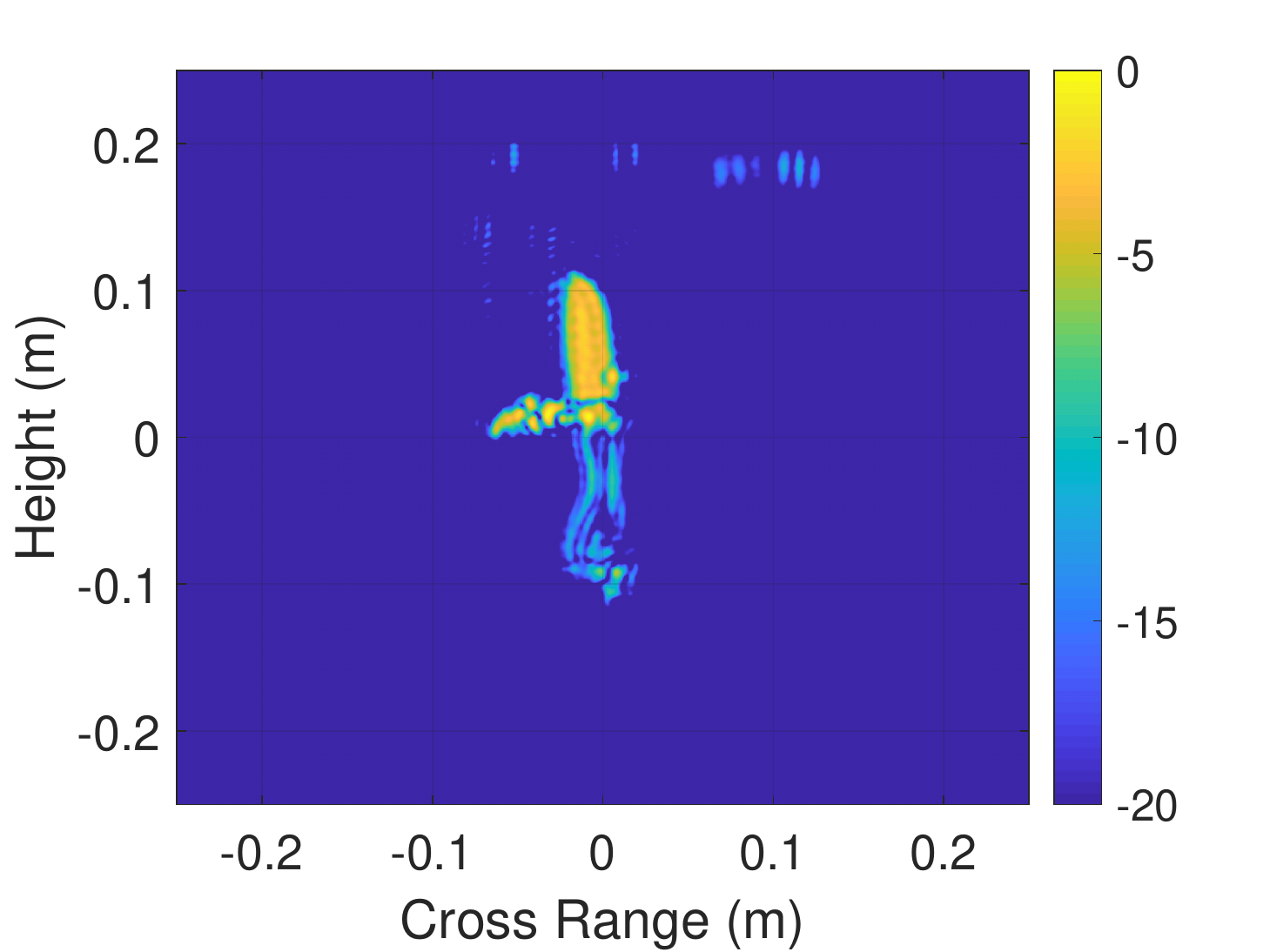}}
		\hfill
		\subfloat[]{\label{b}
			\includegraphics[width=1.69in]{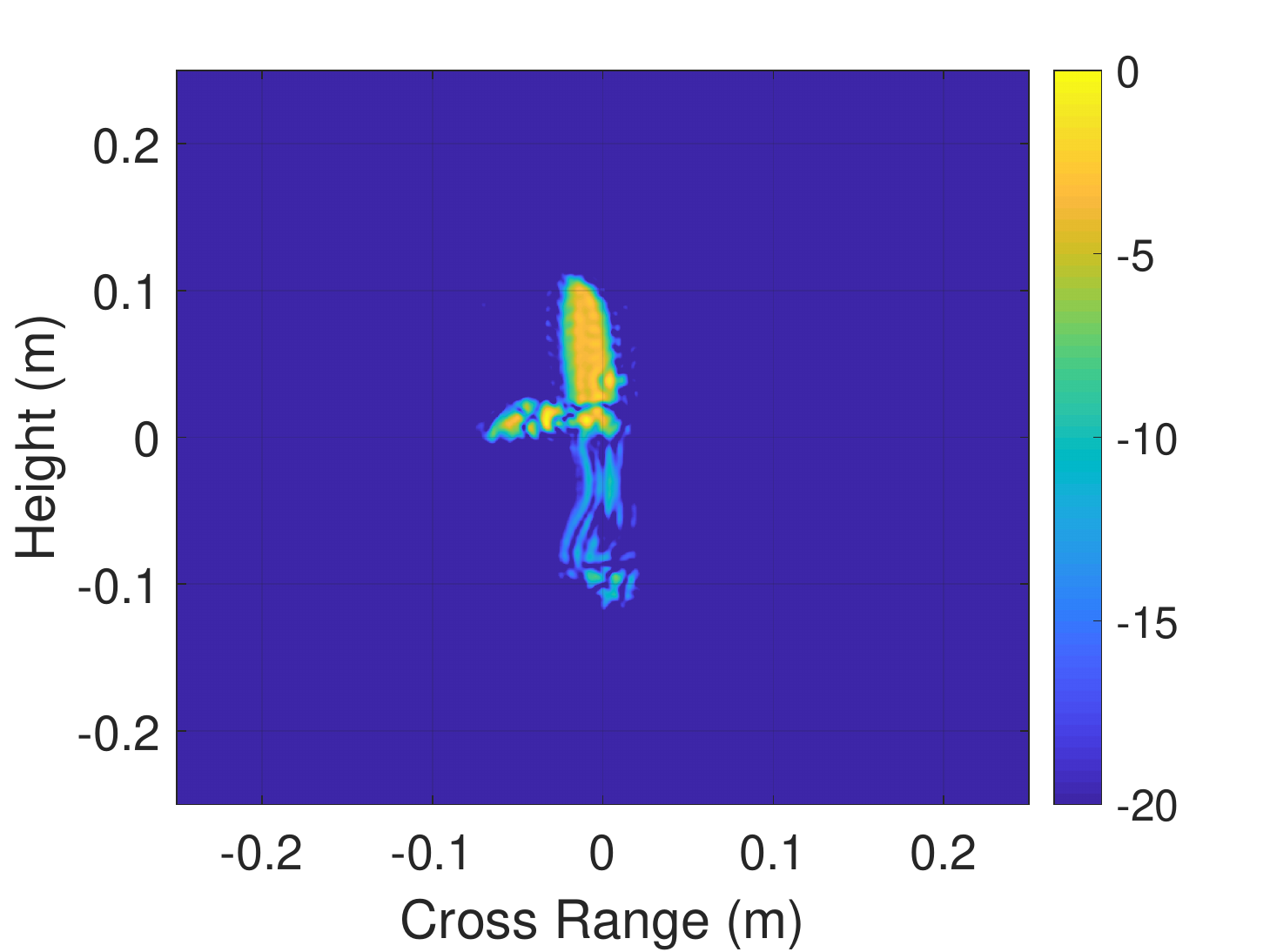}}
		\hfill
		\centering
		\subfloat[]{\label{c}
			\includegraphics[width=1.69in]{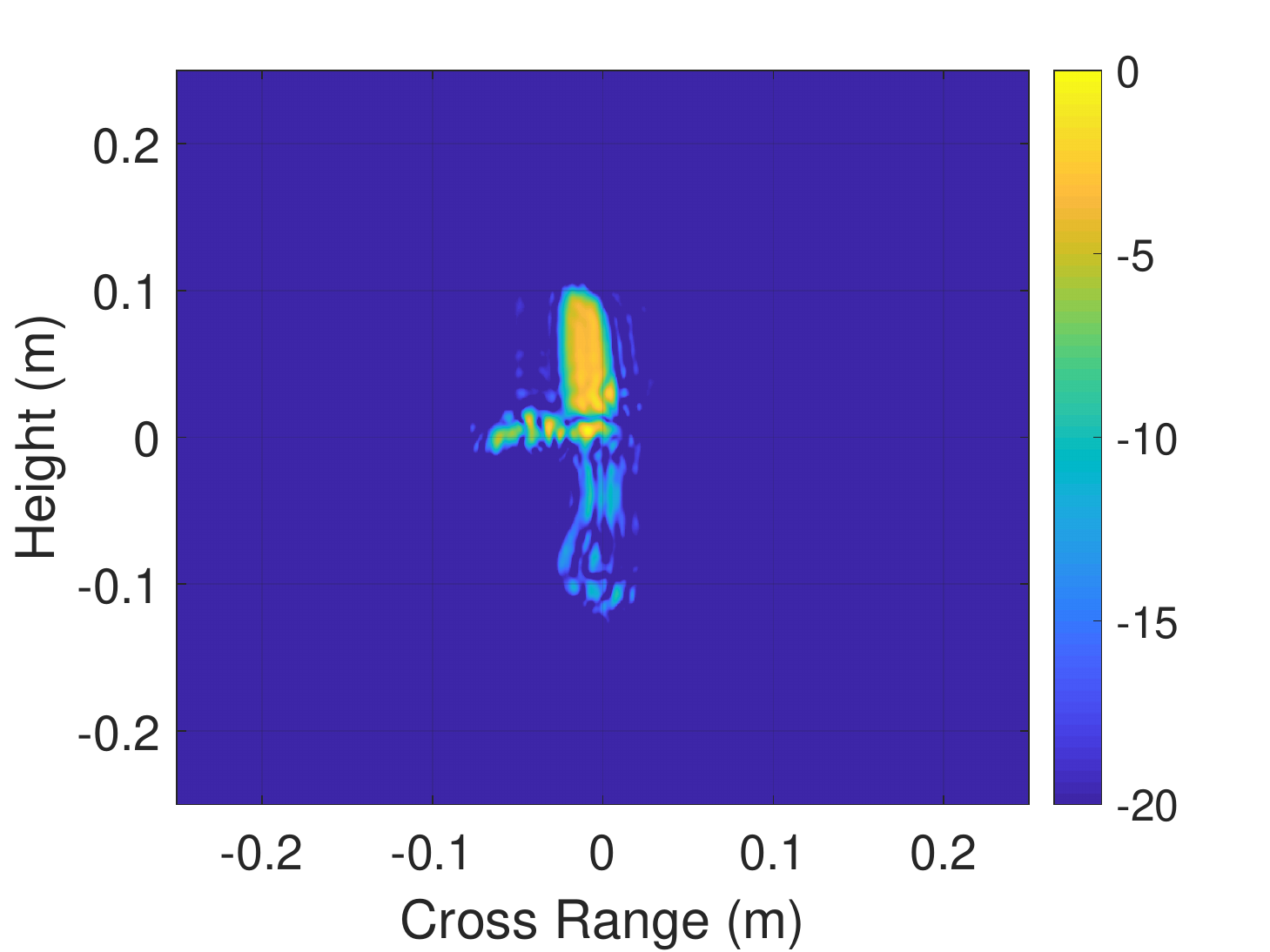}}
		\hfill
		\subfloat[]{\label{d}
			\includegraphics[width=1.69in]{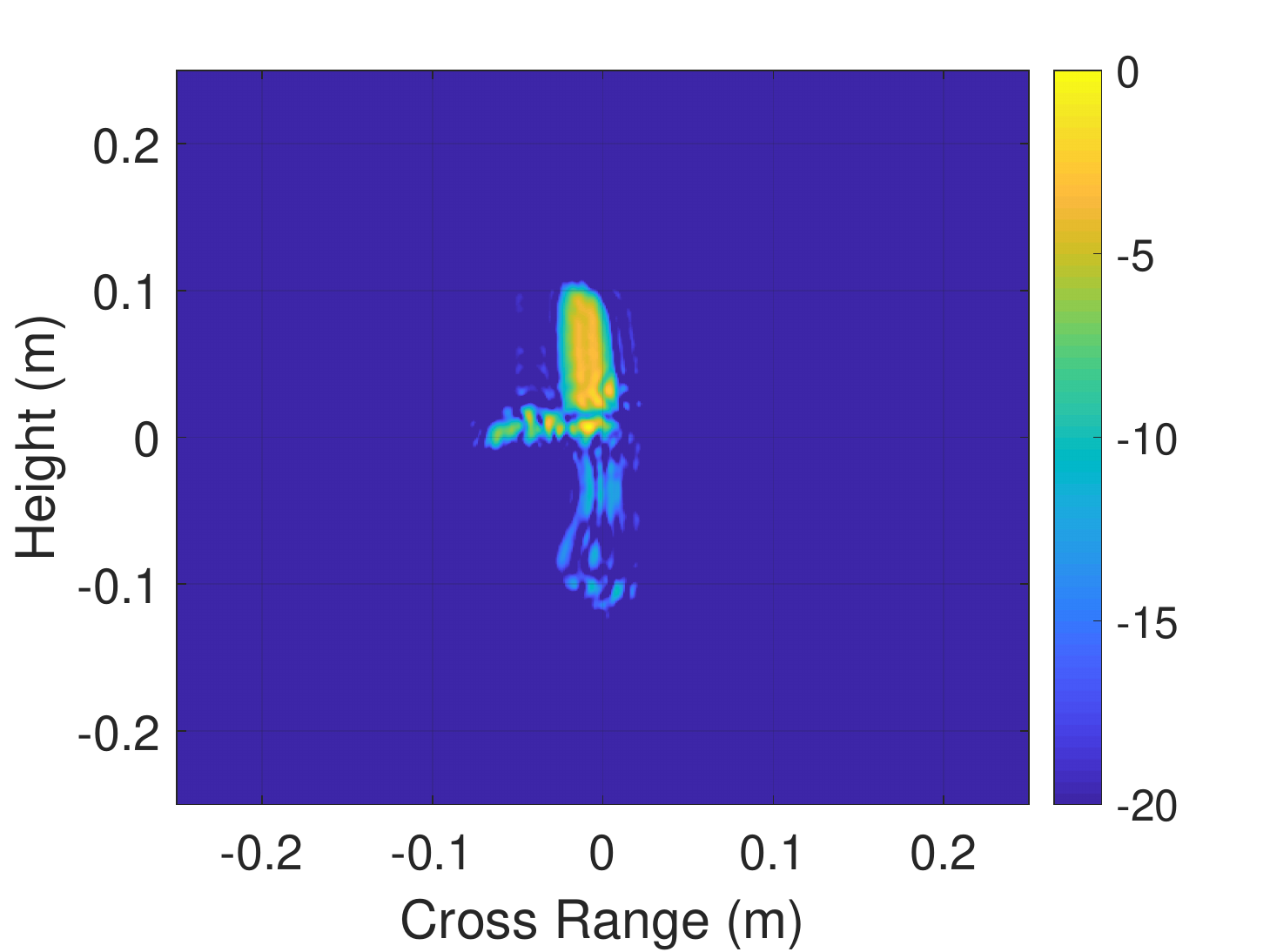}}
		\hfill
		\\	
		\caption{Imaging results  via  monostatic array by (a) the proposed algorithm, and (b) BP; and via multistatic array by (c) the proposed algorithm, and (d) BP.}
		\label{experimental_results}
	\end{figure}
	Finally, the proposed technique is validated experimentally by a prototype of imaging system, as shown in Fig. \ref{experiment_setup}, consisting of a vector network analyzer (VNA), a rotating platform, and a mechanical scanning platform with two independent scanners. Two antennas are connected to the VNA, one as the transmitter, and the other as the receiver. 
	The monostatic and multistatic configurations of antennas are demonstrated in Figs. \ref{experiment_setup}\subref{a} and \ref{experiment_setup}\subref{b}, respectively.	
	The  target under test is a knife that is fixed to a foam platform,  as shown in Fig. \ref{experiment_setup}\subref{c}, whose orientation can be adjusted by the underneath rotating platform, so as to form an aperture like the one  in Fig. \ref{array_geometry}.
	Due to the limit of scanners, we restrict the transmit-receive antenna pairs occuring only within each section (each straight part) of the polyline array, which has no virtual difference between this one and the array described in Section II. B since we utilize time-domain processing along the array dimension.  
	The measurement distance is set to be 0.7 m. Other parameters are the same with those of simulations. 
	
	The 2-D imaging results corresponding to the horizontal-elevation dimension are shown in Fig. \ref{experimental_results}.
	It is noted that comparable imaging performance can be obtained by using the proposed algorithms and BP for both the monostatic and multistatic polyline arrays, which further indicates the effectiveness of the proposed array topology and the corresponding imaging algorithms.
	
	

	\section{Conclusions}
	We presented a polyline  array-based imaging scheme, associated with mechanical scanning along its perpendicular direction, for near-field 3-D MMW imaging. The polyline array can provide better  observation angles than the traditional linear or planar arrays, which is important for the safety inspection of human body. On the other hand, the polyline array is much easier to be fabricated than a circular-arc array.
	The polyline array can be realized as either a monostatic array or a multistatic one.
	The corresponding imaging algorithms were also proposed through using a hybrid processing in the time domain and the spatial frequency domain, accompanied by NUFFT, to improve the computational efficiency.  
	The imaging performance was demonstrated through numerical simulations and experimental results. 


	\ifCLASSOPTIONcaptionsoff
	\newpage
	\fi

	\bibliography{IEEEabrv,full}
	\bibliographystyle{IEEEtran}
	

\end{document}